\documentclass[12pt,preprint,epsf,epsfig,graphics]{aastex}

\shorttitle{Short- and Long-term Chromospheric Activity}
\shortauthors{Shkolnik, Walker, Bohlender, Gu, K\"urster}

\begin{document}


\title{Hot Jupiters and Hot Spots: The Short- and Long-term Chromospheric Activity on Stars with Giant Planets\footnote{Based on observations collected at the Canada-France-Hawaii Telescope operated by the National Research Council of Canada, the Centre
National de la Recherche Scientifique of France, and the University of
Hawaii as well as data from the European Southern Observatory's Very Large Telescope,
    Chile (programme ESO No. 73.C-0694).} }


\author{E.~Shkolnik and G.A.H.~Walker}
\affil{Department of Physics \& Astronomy, University of British Columbia,
    6224 Agricultural Rd., Vancouver  BC, Canada V6T 1Z1}
\email{shkolnik@physics.ubc.ca}
\email{gordonwa@uvic.ca}

\author{D.A. Bohlender}
\affil{Herzberg Institute for Astrophysics, National Research Council of Canada\\
Victoria BC, Canada V9E 2E7}
\email{david.bohlender@nrc-cnrc.gc.ca}

\author{P.-G. Gu\altaffilmark{}}
\affil{Institute of Astronomy \& Astrophysics, Academia Sinica, Taipei, Taiwan 11529}
\email{gu@asiaa.sinica.edu.tw}

\and

\author{M. K\"urster\altaffilmark{}}
\affil{Max-Planck-Institut f\"ur Astronomie,
K\"onigstuhl 17, D-69117 Heidelberg, Germany}
\email{kuerster@mpia-hd.mpg.de}

\begin{abstract}

We monitored the chromospheric activity in the Ca II H \&
K lines of 13 solar-type stars (including the Sun); 8 of them over three
years at the CFHT and 5 in a single run at the VLT. Ten of the 13 targets
have close planetary companions. All of the stars observed at the CFHT
show long-term (months to years) changes in H \& K intensity levels. Four
stars display short-term (days) cyclical activity. For two, HD~73256 and
$\kappa^{1}$~Ceti, the activity is likely associated with an active
region rotating with the star, however, the flaring in excess of the
rotational modulation may be associated with a hot jupiter. A planetary companion remains
a possibility for $\kappa^{1}$~Ceti. For the other two,
HD~179949 and $\upsilon$ And, the cyclic variation is synchronized to the
hot jupiter's orbit. For both stars this synchronicity with the orbit is clearly seen in
two out of three epochs. The effect is only marginal in the third epoch at which the
seasonal level of chromospheric activity had changed for both stars.
Short-term chromospheric activity appears weakly dependent on the mean
K-line reversal intensities for the sample of 13
stars. Also, a suggestive correlation exists between this activity and the $M_{p}$sin$i$ of
the star's hot jupiter. Because of their small separation ($\leq$ 0.1
AU), many of the hot jupiters lie within the Alfv\'en radius of their
host stars which allows a direct magnetic interaction with the stellar
surface. We discuss the conditions under which a planet's
magnetic field might induce activity on the stellar surface and why no
such effect was seen for the prime candidate, $\tau$ Boo. This work
opens up the possibility of characterizing
planet-star interactions, with implications for extrasolar planet
magnetic fields and the energy contribution to stellar atmospheres.

\end{abstract}




\section{Introduction}\label{intro}

In the past decade, $\approx$~130 extrasolar planets have been discovered as companions to late-type dwarfs mostly from Precise Radial Velocities (PRV; e.g.~Walker et al.~1995) from which the Keplerian orbital parameters and minimum planetary masses can be derived.  From transiting planets, of which five are known, true planetary masses, radii, densities and albedo-dependent estimates of surface temperature are available. To date, a handful of experiments have lead to further information about these planets, including Lyman~$\alpha$ (Vidal-Madjar et al.~2003) and sodium detections (Charbonneau et al.~2002) in the atmosphere of the transiting planet around HD~209458.  A current review of follow-up techniques to probe planetary characteristics has been presented by Charbonneau (2003).

The presence of a giant planet likely influences its parent star beyond the dynamical perturbations measured by the PRV method. Robinson \& Bopp (1987) caught the signatures of ``superflares" with energies of $\sim$~10$^{35}$ ergs (Schaefer et al.~2000) on nine solar analogs which have no otherwise unusual properties such as very rapid rotation or very high chromospheric activity.  Rubenstein \& Schaefer (2000) suggested that these anomalous superflares were stimulated by an unseen close-in extrasolar giant planet (CEGP, also known as a `hot jupiter').  They explored the possibility of magnetic reconnection events occurring between both the star's and the planet's entangled magnetic fields.

Cuntz et al. (2000) suggested a more consistent observable interaction between a
parent star and its hot jupiter in the form of enhanced stellar activity
 of the star's outer atmosphere. This interaction can take the form of tidal and/or magnetic heating of the plasma. Both the star and its planet experience strong tidal forces as well as repeated magnetic reconnection with a large planetary magnetosphere. If such planet-induced heating of the star is confined to a narrow range in stellar longitude, the heated regions likely track the planet as it orbits. This implies that the period of any observed activity would be correlated with the planet's orbital period $P_{orb}$ such that tidally induced activity has a period of $\sim P_{orb}/2$ and magnetic activity, a period of $P_{orb}$. In the simplest configuration of magnetic interaction, the expected enhancement would occur near the sub-planetary point, which defines orbital phase $\phi = 0$ at the time when the planet is in front of the star relative to the line-of-sight.

A partial analogy can be made to the Jupiter-Io system (Zarka et al.~2001) where the volcanically
active moon of Jupiter, orbiting at a distance of 5.9 $R_{J}$, constantly
couples with Jupiter's magnetosphere, leaving two footprints at high positive and negative latitudes
on the planet's surface. Plasma flows along the magnetic field lines, making up the Io Flux Tube, as these footprints follow Io in its orbit. Even though the analogy is limited, a similar phenomenon may occur between CEGPs and their stars such that coupling between the magnetic fields of the planet and the star may cause footprints or ``hotspots" on the star's surface which follow the planet and have a period close to the planet's orbital period Similarly, auroral emission may be stimulated on the planet's atmosphere (Zarka et al.~2001) but no searches for planetary radio emissions have yet been successful (e.g. Bastian et al.~2000, Farrell et al.~2003, Ryabov et al.~2003).

More recently, Santos et al.~(2003) observed cyclic photometric variations with a period
very similar to that of the radial velocity (RV) curve for the K dwarf HD~192263. The
stability of the 24.4-day periodic radial velocity through almost 4 years of data (Santos et al.~2000, Henry et al.~2002) rules out the interpretation that stellar activity alone is the
cause of the RV curve and supports the existence of a planetary-sized
companion around the star. However, they question what might cause a quasi-stable photometric period that coincides with the planetary orbit.  They offered planet-induced magnetic activity off-set by 90$^{\circ}$ from the sub-planetary point as an explanation.

Even though the interpretations of Rubenstein \& Schaefer (2000) and Santos et al. (2003) are uncertain, there exists ample observational evidence of such tidal and magnetic interactions in the exaggerated case of the RS Canum Venaticorum (RS CVn) stars, which are tightly-orbiting binary systems consisting of two chromospherically active late-type stars. For example, Catalano et al.~(1996) found as many starspots and plages within 45$^{\circ}$ of the sub-binary point\footnote{The `sub-binary' point refers to the longitude on the star that faces the binary system's center-of-mass.} as on the rest of the stellar surface for several RS CVn systems. $\lambda$~And, a relatively long-period system ($P_{orb}$ = 20.1 d; Walker 1944), shows modulation of the Mg~II UV chromospheric emission lines with a period of 10 days, half the orbital period. Glebocki et al.~(1986) interpreted this as a tidal heating of the primary by its companion, possibly a brown dwarf (Donati et al.~1995). And, in our own Ca II H \& K observations of ER~Vul, an RS CVn system with two G~V dwarfs and $P_{orb}$ = 0.69 d, we see clear enhancements near the sub-binary longitudes (Shkolnik et al.~2004).

With these scenarios in mind, we searched for
periodic chromospheric heating by monitoring the Ca~II H \& K emission in
stars with giant planets within a few stellar radii. We chose to study the tightest observable systems since the tidal and magnetic interactions depend on the distance from the planet to the star as $1/r^{3}$ and $1/r^{2}$, respectively. Of the known extrasolar planets, about 20\% have semi-major axes of less than 0.1 AU and masses comparable to Jupiter's (Schneider 2004). It is expected that these planets have magnetic fields also similar to Jupiter's (4.3 G).\footnote{However, if tidally locked, the planets' rotation rates may be much lower than Jupiter's possibly causing their magnetic fields to be substantially smaller (e.g.~Sanch\'ez-Lavega 2004).}  It is also reasonable to assume that any magnetic interaction would be greatest in the outermost layers of the star, namely the chromosphere, transition region and the corona due to their proximity to the planet, low density, and nonradiative heat sources. The broad, deep photospheric absorption of the Ca~II H \& K lines allows the chromospheric emission to be seen at higher contrast. Because of this and the accessibility from the ground, the H \& K reversals are an optimal choice with which to monitor chromospheric heating of these sun-like stars.

Our program stars have orbital periods between 2.5 and 4.6 days, eccentricities $\approx$~0 and semi-major axes $< 0.06$ AU.  These systems offer the best chance of observing upper atmospheric heating.  The first five systems we observed were $\tau$~Boo, HD~179949, HD~209458, 51~Peg and $\upsilon$~And from the Canada-France-Hawaii Telescope (CFHT).  The first results from 2001 and 2002 observations, including the first evidence of planet-induced magnetic heating of HD~179949, were published in Shkolnik et al.~(2003).  We later extended the experiment at the Very Large Telescope (VLT) to include five southern targets, HD~46357, HD~73256, HD~75289, HD~76700, and HD~83443.  The system parameters for the program stars are listed in Table 1 along with our two standards, $\tau$~Ceti and the Sun.

In this paper, we compile our 2003 CFHT data with those of previous years and include the recent VLT observations. A broader understanding of stellar activity, its cycles, and planet-induced chromospheric heating emerges. We also observed $\kappa^{1}$~Ceti, a young (650 $-$ 750 Myr old), chromospherically active, solar analog. It was one of the nine stars reported to have anomalous superflare activity by Robinson \& Bopp (1987) possibly caused by magnetic interactions with a close-in giant planet (Rubenstein \& Schaefer 2000). Of their sample, only $\kappa^{1}$~Ceti has been looked at by the PRV planet searches of Walker et al.~(1995), Cumming et al.~(1999) and Halbwachs et al.~(2003); none of which has detected a planet.  Interestingly though, Walker et al.~(1995) observed a rapid RV change of 80 m~s$^{-1}$ in 1988 which could have been caused by a planet in highly elliptical orbit. However, it coincided with an equally sharp increase in the Ca II chromospheric activity indicator at 8662 \AA\/ implying the RV jump was likely due to changes intrinsic to the star.  We may have observed this same phenomenon in our Ca II data.

The details of our CFHT and VLT observations are outlined in Section~\ref{spectra}. We briefly discuss the precise differential radial velocities which yielded updated ephemerides for the planetary systems such that orbital phases could be determined to better than 0.02. In Section~\ref{caII} we discuss our analysis and results of the Ca II H \& K measurements including long-term, short-term and rotational modulation.  A theoretical discussion of the physical requirements for magnetic interactions between stars and their hot jupiters is presented in Section~\ref{theory} with future experiments suggested in Section~\ref{summary}.

\begin{deluxetable}{ccccccccccl}

\tabletypesize{\footnotesize}
\tablecaption{Stellar and Orbital Parameters \label{starpars}}
\tablewidth{0pt}
\tablehead{
\colhead{~ Star} &
\colhead{Spectral} &
\colhead{$v$sin$i$} &
\colhead{$P_{rot}$} &
\colhead{$P_{orb}$\tablenotemark{a,\rm b}} &
\colhead{$M_{p}$sin$i$\tablenotemark{b}} &
\colhead{Semi-major\tablenotemark{b}} &
\colhead{$\langle$K$\rangle$\tablenotemark{c}} &
\colhead{$\langle$K$\arcmin\rangle$\tablenotemark{d}} &
\colhead{$\langle$MAD K$\rangle$\tablenotemark{e}} &
\\
\colhead{} &
\colhead{Type} &
\colhead{(km s$^{-1}$)} &
\colhead{(days)} &
\colhead{(days)} &
\colhead{($M_{J}$)} &
\colhead{axis (AU)} &
\colhead{(\AA)} &
\colhead{(\AA)} &
\colhead{(\AA)} &
}

\startdata

$\tau$~Boo  &    F7~IV &  14.8$\pm$0.3  &   3.2\tablenotemark{f}   &3.313  &3.87   & 0.046 &  0.323 & 0.177 &  0.0020\\
HD~179949   &    F8~V  &  6.3$\pm$0.9   &  $<$9\tablenotemark{g}   &3.092  &0.98   & 0.045 &  0.399 & 0.202&  0.0051\\
HD~209458   &    G0~V  &  4.2$\pm$0.5   &  16\tablenotemark{h}  & 3.525  &0.69\tablenotemark{i}   & 0.045 &  0.192 & 0.077&  0.0009\\
51~Peg  &	G2~IV  & 2.4$\pm$0.3    & 21.9\tablenotemark{f}   &4.231  &0.47    &0.05   & 0.178  & 0.071 & 0.0008\\
$\upsilon$~And & F7~V  &  9.0$\pm$0.4   &  14\tablenotemark{f}    & 4.618  &0.71\tablenotemark{j}   & 0.059 &  0.254 & 0.091&  0.0016\\
HD~46375    &    K1~IV &  $<$2  	&43\tablenotemark{k}      &3.024   &0.25   &0.041  & 0.339  & 0.233 & 0.0011\tablenotemark{l}\\
HD~73256    &    G8/K0~V & 3.22$\pm$0.32&   14\tablenotemark{m}  &    2.549 &1.85 &   0.037 &  0.995 & 0.899 &  0.0051\tablenotemark{n}\\
HD~75289    &    G0~V  &  4.37    	&15.95\tablenotemark{m} &  3.510  &0.42  &  0.046 &  0.177 & 0.062  & 0.0003\\
HD~76700    &    G6V   &  $<$2  & ---   &         3.971 &	0.20  & 0.049 &  0.212 & 0.095 &  0.0004\\
HD~83443    &    K0V   &  1.4    & 	35\tablenotemark{k}  &  2.986 &0.38 &   0.039 &   0.314  & 0.211 & 0.0010\\
$\kappa^{1}$~Cet & G5~V &    4.64$\pm$0.11  & 9.3\tablenotemark{o} &    ---   &  ---  &   ---  &   0.906 & 0.815 &  0.0059\tablenotemark{n}\\
$\tau$~Cet   &   G8~V  &  1      & 	33   &   ---  &   ---   &  --- &    0.201  & 0.102 & 0.0003\\
Sun     &	G2~V  &  1.73$\pm$0.3 &   27  &    --- &    --- &    ---  &   0.298 & 0.137 &  0.0005\\

\enddata

\tablenotetext{a}{The first five periods are from this work. (See Section \ref{deltaRVs}.) See Note b for the last five.}
\tablenotetext{b}{Published orbital solutions: $\tau$ Boo \& $\upsilon$ And $-$ Butler
et al.~1997; HD 179949 $-$ Tinney et al.~2001: 51 Peg $-$ Marcy et al.~1996; HD 209458 $-$ Charbonneau et al.~1999); HD 46375 $-$ Marcy et al.~2000; HD 73256 $-$ Udry et al.~2003; HD 75289 $-$ Udry et al.~2000; HD 76700 $-$ Tinney et al.~2002; HD 83443 $-$ Mayor et al.~2004.}
\tablenotetext{c}{Total integrated intensity bounded by the K1 features of the mean normalized Ca II K core. These values are relative to the normalization points near 3930 and 3937 \AA\/ which are at approximately 1/3 of the continuum at 3950 \AA.}
\tablenotetext{d}{We subtracted the photospheric emission from $\langle$K$\rangle$ in order to measure the mean integrated chromospheric emission $\langle$K$\arcmin\rangle$ using data from Wright et al.~(2004). (See text for more details.)}
\tablenotetext{e}{Average integrated `intensity' of the mean absolute deviation (MAD) of the K residuals, per observing run}
\tablenotetext{f}{Henry et al.~2000}
\tablenotetext{g}{$P_{rot}$ is calculated from the $v$sin$i$ and a stellar radius of 1.24 R$_{\odot}$ (Tinney et al. 2001).}
\tablenotetext{h}{Mazeh et al.~2000}
\tablenotetext{i}{Transiting system; $i = 86.1^{\circ}\pm 0.1^{\circ}$ (Mazeh et al. 2000)}
\tablenotetext{j}{Closest of three known planets in the system}
\tablenotetext{k}{Derived from empirical fits (Noyes et al.~1984) of the Ca II $R'_{HK}$ index (Wright et al.~2004)}
\tablenotetext{l}{$\langle$MADK$\rangle$ was corrected for a $\approx$~30\% variable contribution by telluric Ca II emission since it was the first star observed after sunset.}
\tablenotetext{m}{Udry et al.~2003}
\tablenotetext{n}{Value corrected to remove modulation due to rotation.  For HD 73256, the non-corrected value is 0.0155 and for $\kappa^1$ Ceti, 0.0182.}
\tablenotetext{0}{Rucinski et al.~2004}

\end{deluxetable}

\section{The Spectra}\label{spectra}

\subsection{CFHT Observations}\label{cfht_obs}

The observations were made with the 3.6-m Canada-France-Hawaii Telescope (CFHT) on 3.5, 4, 2, and 5 nights in
2001 August, 2002 July, 2002 August, and 2003 September, respectively. We
used the Gecko \'{e}chellette spectrograph fiber fed by CAFE (CAssegrain
Fiber Environment) from the Cassegrain to Coud\'{e} focus (Baudrand \&
Vitry 2000). Spectra were centered at 3947 \AA\/ which was isolated by a
UV grism (300 lines mm$^{-1}$) with $\simeq$~60~\AA\/ intercepted by the
CCD. The dispersion was 0.0136 \AA\/ pixel$^{-1}$ and the 2.64-pixel FWHM
of the thorium-argon (Th/Ar) lines corresponded to a spectral resolution
of R = 110,000. The detector was a back-illuminated EEV CCD (13.5
$\mu$m$^{2}$ pixels, 200 $\times$ 4500 pixels) with spectral dispersion
along the rows of the device.

\begin{figure}
\epsscale{.80}
\plotone{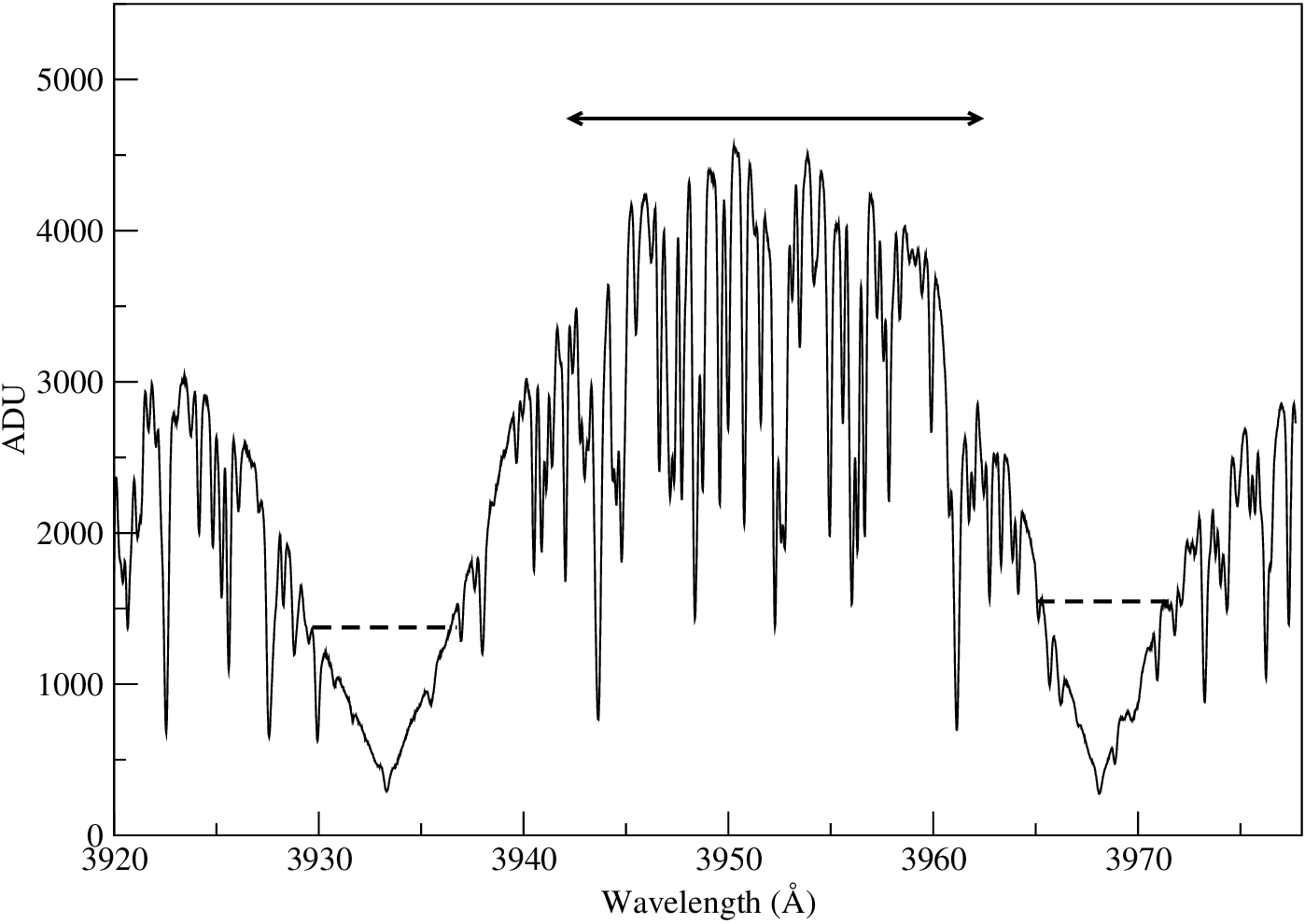}
\caption{A single spectrum of $\upsilon$~And. The arrow defines the 3942 $-$ 3963 \AA\/ region used to measure the differential radial
velocities. The dashed lines show the levels of normalization
use to isolate the H and K cores.
\label{upsAnd_spec}}
\end{figure}

To remove the baseline from each observation, the appropriate mean darks
were subtracted from all the exposures. Flat-fields were then normalized
to a mean value of unity along each row. All the stellar exposures
observed on a given night were combined into a mean spectrum to define a
single aperture for the extraction of all stellar and comparison
exposures, including subtraction of residual background between spectral
orders (prior to flat-fielding). This aperture was ultimately used to
extract one-dimensional spectra of the individual stellar and comparison
exposures and of the mean, normalized flat-field. This one-dimensional flat used to obtain the most consistent flat-fielded spectra possible.
All the data were processed
with standard IRAF (Image Reduction and Analysis
Facility) routines.\footnote[1]{IRAF is distributed by the National Optical
Astronomy Observatories, which is operated by the Association of
Universities for Research in Astronomy, Inc.~(AURA) under cooperative
agreement with the National Science Foundation.}
Wavelength calibration was done using the Th/Ar arcs taken before and after each
spectrum. We required frequent arcs in order to track the CCD drift
or any creaking in the system throughout the night. Heliocentric and
differential radial velocity corrections were applied to each stellar
spectrum using IRAF's {\it rvcorrect} routine. A specimen, flat-fielded
spectrum of $\upsilon$~And is shown in Figure~\ref{upsAnd_spec}. The Ca~II H
\& K reversals are weak yet visible at 3968 and 3933 \AA\/.  The final spectra were of high S/N reaching $\approx$ 500 per pixel (or 4300 \AA$^{-1}$) in the continuum and 150 (or 1290 \AA$^{-1}$) in the H \& K cores.

Spectra with comparable S/N were taken of two stars known not to have close-in giant planets, $\tau$ Ceti and the Sun (sky spectra were taken at dusk). Table 2 of Walker et al. (2003a) lists the five CFHT program stars plus $\tau$~Ceti, including their U magnitudes, exposure times, and typical S/N.

\subsubsection{Precise Differential Radial Velocities}\label{deltaRVs}

Differential radial velocities ($\Delta$RVs) were estimated with the {\it fxcor} routine in IRAF (version PC-IRAF V2.12) which performs a Fourier cross-correlation on dispersion-corrected spectra. We used the first spectrum in the series for each star as the template, hence, all $\Delta$RVs are relative to the first spectrum on the first night of the run. Both the template and the input spectrum were normalized
with a low order polynomial. The correlation used only the $\sim$~20 \AA\/ region between the H \& K lines, that part of the spectrum bounded by (and including) the two strong Al~I lines (3942 $-$ 3963 \AA) as shown in Figure~\ref{upsAnd_spec}.

\begin{figure}
\epsscale{.70}
\plotone{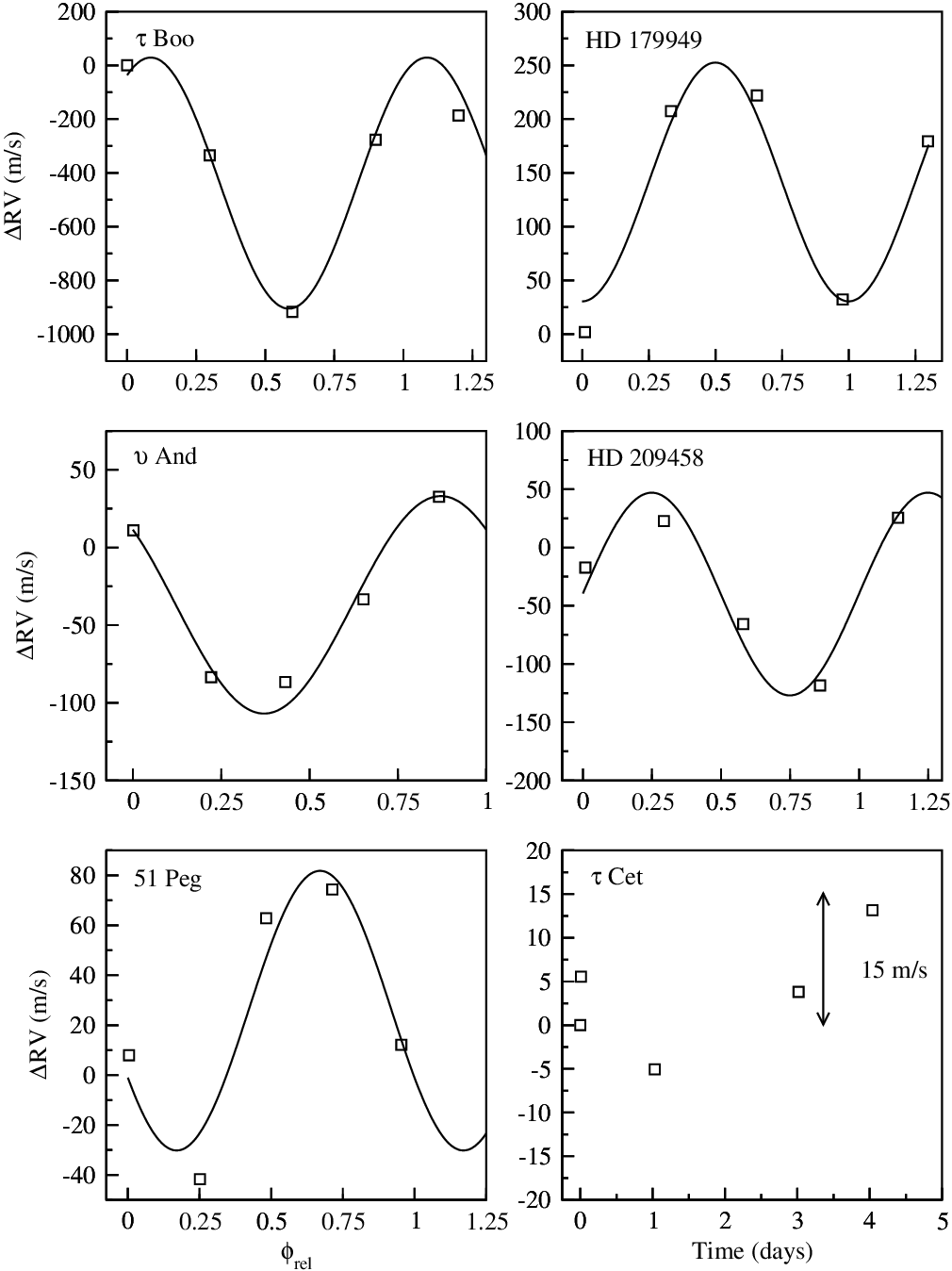}
\caption{The differential radial velocities for the five `51 Peg' stars measured in 2003 September are plotted as a function of relative orbital phase, and the unvarying star, $\tau$~Ceti, as a function of time. Each sine curve has the  planetary orbital period from Table 1 and published amplitude for the star and has been shifted in phase to give the best fit to the $\Delta$RVs. The measurement error of the individual points is $<$~5~m~s$^{-1}$. The $\sigma_{RV}$ of the $\Delta$RV are listed in Table~\ref{ephemerides}. Note that $\tau$~Ceti's $\Delta$RVs are plotted on a much smaller scale than the others and have a $\sigma_{RV}$ of 7 m~s$^{-1}$, making the average $\sigma_{RV}$ for all the data 17 m~s$^{-1}$.
\label{rv2003}}
\end{figure}

The $\Delta$RVs measured during the 2002 July run for the five `51 Peg' stars can be found in Walker et al.~(2003a). The $\Delta$RVs from 2003 September are plotted in Figure~\ref{rv2003}. After the orbit is removed, the average $\sigma_{RV}$ is 17 m~s$^{-1}$. The two stars observed at the highest S/N have $\sigma_{RV}$ of 7 and 9 m~s$^{-1}$. This precision may be the best achievable for a single spectrum with this spectrograph.  The excellent $\Delta$RVs
yield current orbital ephemerides and hence accurate phases ($\pm~0.02$) for each observation. Using the ephemerides of the planets' discovery orbits, we tabulated the 2003 September times of sub-planetary position ($\phi$ = 0) with revised orbital periods. These are listed in Table~\ref{ephemerides}.


\begin{deluxetable}{ccccccl}
\tabletypesize{\footnotesize}
\tablecaption{2003 September Ephemerides \label{ephemerides} }
\tablewidth{0pt}
\tablehead{
\colhead{~ Star} &
\colhead{$\sigma_{\Delta RV}$} &
\colhead{HJD at $\phi=0$} &
\colhead{$\delta$(HJD)\tablenotemark{a}} &
\colhead{Revised $P_{orb}$} &
\colhead{$\delta(P_{orb})$\tablenotemark{a}} &
\\
\colhead{} &
\colhead{m~s$^{-1}$} &
\colhead{days} &
\colhead{days} &
\colhead{days} &
\colhead{days} &
}

\startdata

$\tau$~Boo  &    33  &    2452892.864  &   0.066 &  3.31250 & 0.00026 \\
HD~179949   &    19  &    2452894.114   &  0.062 &  3.09246 & 0.00031\\
HD~209458   &    17  &    2452893.653   &  0.070 &  3.52490 & 0.00020\\
51~Peg & 	 15  &    2452895.868  &   0.085 &  4.23092 & 0.00014\\
$\upsilon$~And &  9    &    2452892.615   &  0.092 &  4.61750 & 0.00052\\

\enddata

\tablenotetext{a}{Uncertainties in the respective measurements}
\end{deluxetable}


\begin{deluxetable}{cccccrccl}
\tabletypesize{\footnotesize} \tablecaption{ The VLT Program Stars \label{vltstars}}
\tablewidth{0pt}
\tablehead{
\colhead{~ Star} &
\colhead{U} &
\colhead{B} &
\colhead{Exp. time} &
\colhead{n\tablenotemark{a}} &
\colhead{S/N\tablenotemark{b}} &
\colhead{S/N\tablenotemark{b}} &
\\
\colhead{} &
\colhead{} &
\colhead{} &
\colhead{s} &
\colhead{} &
\colhead{cont} &
\colhead{core} &
}

\startdata

HD~46375     &   9.33  &  8.7  &   300  &   8    &   510  &   150\\
HD~73256     &    --   &  8.86 &   300  &   9 $\times$ 2 &     520   &  150\\
HD~75289     &   7.04  &  6.94 &   120  &   10   &   790  &   230\\
HD~76700     &   9.17  &  8.76 &   300  &   8    &   540  &   160\\
HD~83443     &   9.54  &  9.03 &   300  &   8    &   420  &   120\\

\enddata

\tablenotetext{a}{Number of spectra taken per night. We observed HD~73256 twice per night for a total of 18 exposures.}
\tablenotetext{b}{Nightly average per 0.015 \AA\/ per pixel in the continuum near 3950~\AA\/ and in the Ca II K core.}

\end{deluxetable}

\subsection{VLT Observations}\label{vlt_obs}

We obtained high-resolution spectra through Visitor mode using the VLT's Ultraviolet and Visual Echelle Spectrograph (UVES) mounted on the 8.2-m Kueyen (UT2) over four photometric half-nights (2004 April 4 $-$ 7).  The standard blue arm setting was used, centered on 4370 \AA, giving a wavelength range of 3750 to 4990 \AA.  We used the CD2 cross-disperser grating (660 g~mm$^{-1}$) with a CCD of 2048 $\times$ 3000 pixels of 15~$\mu$m$^{2}$. Image Slicer \#2 with a slit width of 0.44$\arcsec$ resulted in a resolution of R $\approx$ 75,000.

The data was reduced on-site by the UVES data reduction pipeline that uses the ESO-MIDAS software package within the UVES content.  The data processing consisted of standard procedures: bias subtraction, interorder background correction, cosmic ray hit removal, flat-fielding, and wavelength calibration. The data were wavelength calibrated with Th/Ar arcs attached at the beginning and end of each set of 8 to 10 stellar exposures. The exposure times, number of exposures and the S/N for each star are listed in Table~\ref{vltstars}.

\section{A Comparison of Ca II Emission}\label{caII}

\subsection{Extracting the H, K and Al~I Lines}\label{HKlines}

\begin{figure}
\epsscale{.70}
\plotone{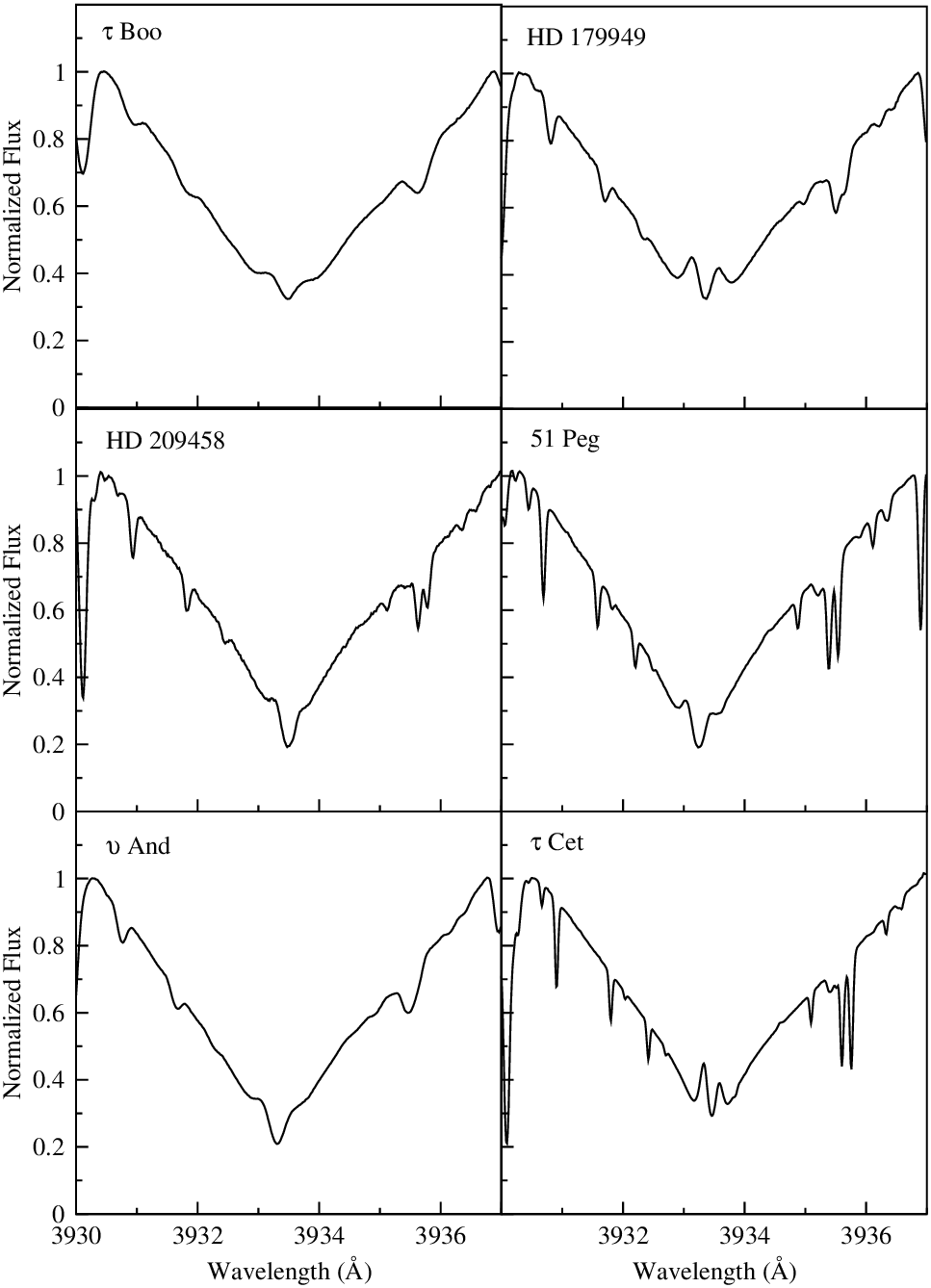}
\caption{The mean normalized Ca II K cores for the CFHT program stars.
\label{cfht_Kcores}}
\end{figure}

\begin{figure}
\epsscale{.70}
\plotone{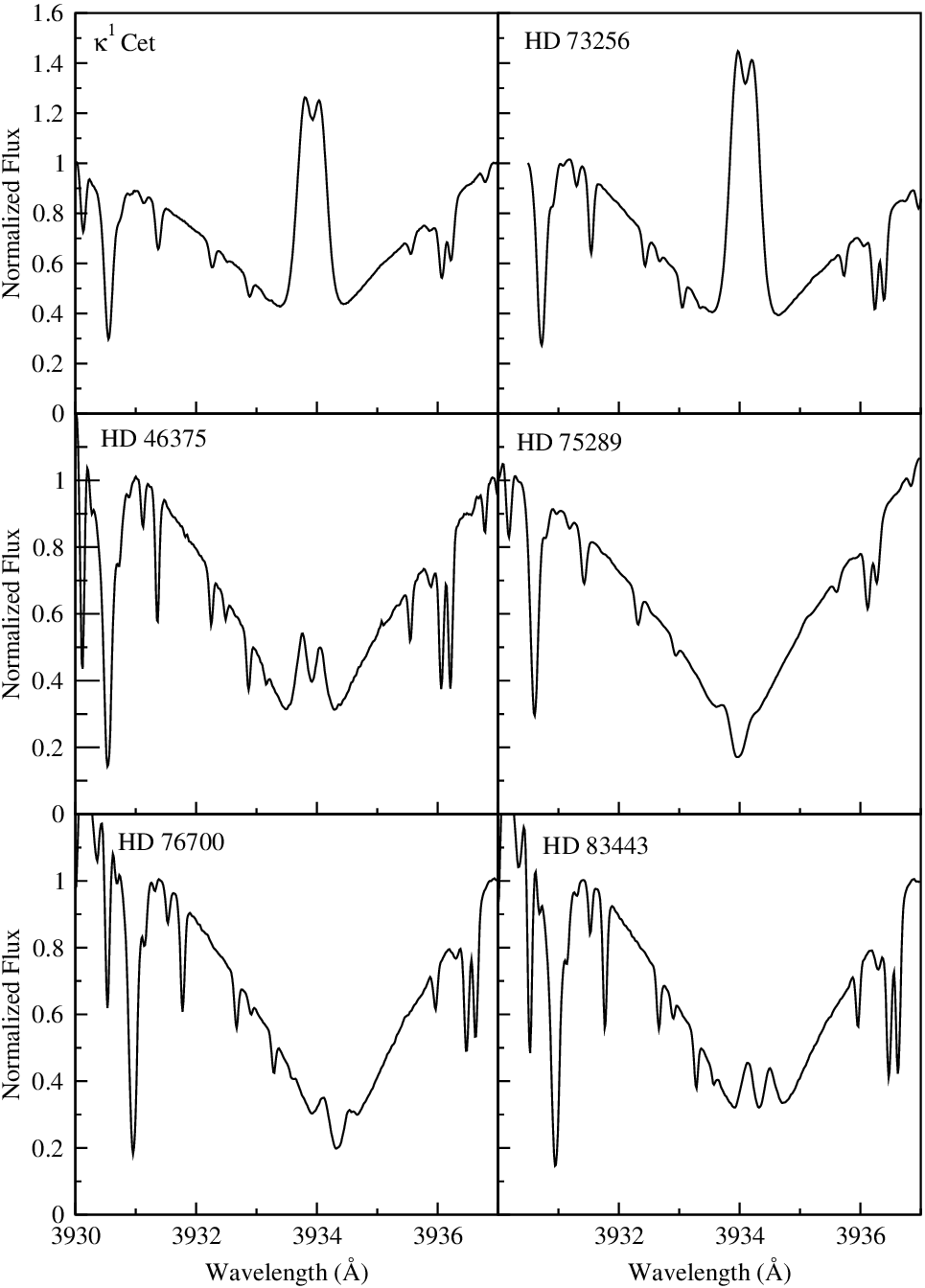}
\caption{The mean normalized Ca II K cores for the VLT program stars plus $\kappa^{1}$~Ceti, which we observed at the CFHT.
\label{vlt_Kcores}}
\end{figure}

The very strong Ca II H and K photospheric lines
suppress the stellar continuum making it difficult to normalize each
60-\AA\/ spectrum consistently.  For this reason, a careful analysis by
which to isolate any modulated Ca II emission was devised. Figure~\ref{upsAnd_spec} shows a flat-fielded spectrum of $\upsilon$~And with the normalization levels marked with dashed lines. The normalization wavelengths were constant for all spectra of a given star.  The 7-\AA\/ spectral ranges, centered on the H and K lines, were chosen to isolate the H and K reversals while minimizing any apparent continuum differences induced by varying illumination of the CCD. This window
is, however, wide enough such that a few photospheric absorption features
appear which could be tested for variability as well. The
mean Ca II K cores for the program stars are shown in Figures~\ref{cfht_Kcores} and \ref{vlt_Kcores}. Also, 7-\AA\/ and 2-\AA\/ cuts, centered on the strong photospheric Al~I
line at 3944 \AA\/ were used as internal standards since the line has comparable depth and S/N to the Ca II lines.

To normalize each sub-spectrum, the end points were set to 1 and
fitted with a straight line. The spectra were grouped by date and a nightly
mean was computed for each of the three lines. The RMS of the Al~I
residuals for each CFHT target star is less than 0.0005 of the normalized mean.  This
is representative of the Al~I residuals for all stars in all four runs. The Al~I line of the VLT data varies with an average RMS of 0.0006. These values demonstrate both the level of stellar photospheric stability as well as
the reliability of the data reduction and analysis.
For all stars observed, both the H \& K emission was non-varying at the 0.001 level on a given night, likely as a result of intra-night (statistical) noise.

\subsection{Long-term Variability}\label{long}

\begin{figure}
\epsscale{.80}
\plotone{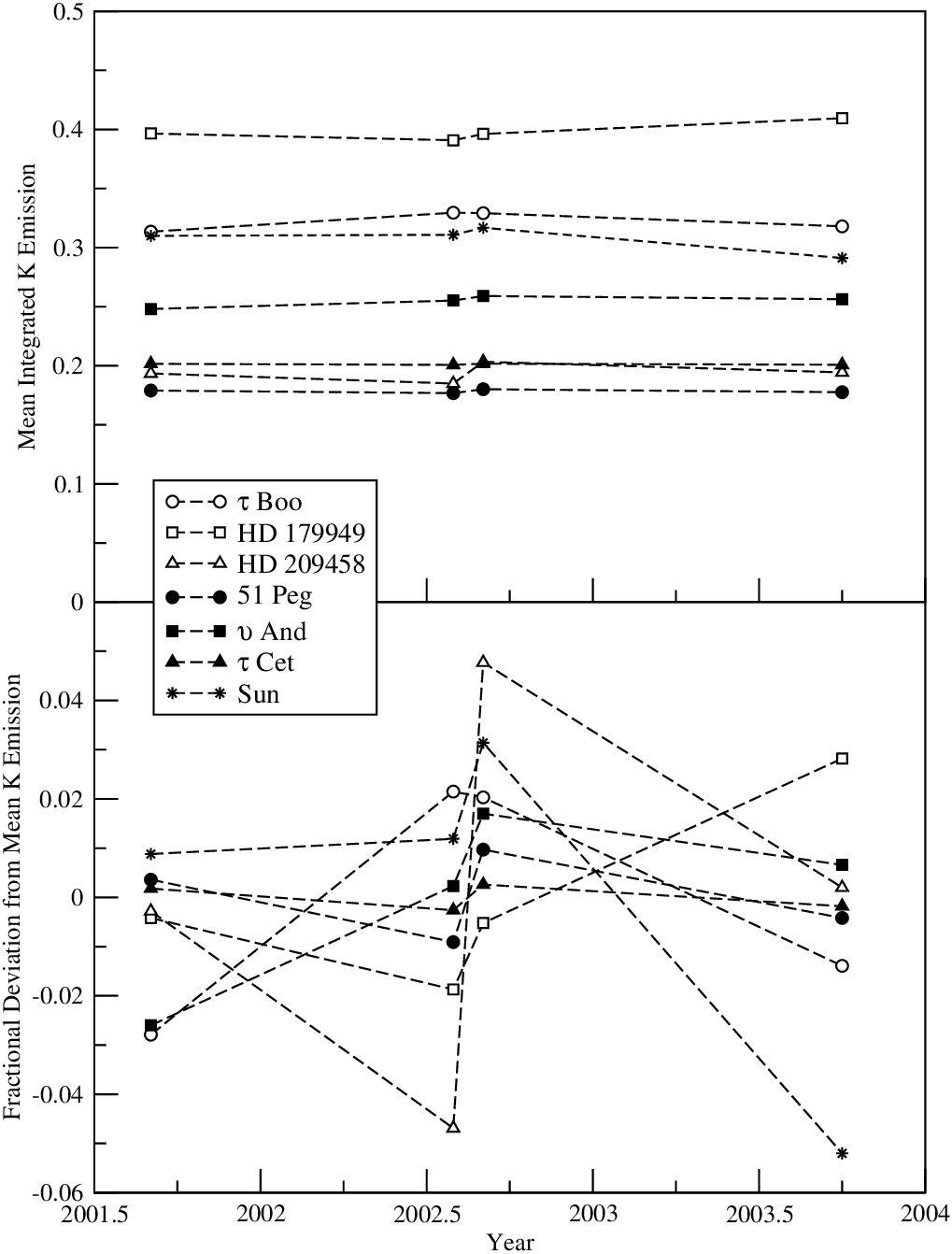}
\caption{Top: the mean integrated Ca II K emission from each of the four runs for the CFHT program stars, including $\tau$~Ceti and the Sun. Units are in equivalent angstroms relative to the normalization points. Bottom: the fractional variation for each star relative to its overall mean K emission.
\label{yearly}}
\end{figure}

With four CFHT observing runs spanning a baseline of over two years, we can compare the long-term variations in the chromospheric levels of the stars. We measure emission strength by integrating across the normalized K cores bounded by the K1 features (Montes et al.~1994). Each star's average integrated K emission $\langle$K$\rangle$ for each observing run is plotted in Figure~\ref{yearly} along with its fractional variation relative to the overall average emission.  This is a good start to tracking the intrinsic stellar activity cycles of these stars. In the case of the Sun, we see the decrease from 2001 August to 2003 September as it declines from solar maximum. However, we also observed the naked-eye sunspot grouping of 2002 August that appeared in our data as a $\approx$~2\% increase in the Ca II emission relative to the other years. Since the variability from run-to-run may also be an indication of active regions on the disk of the star, we require more frequent monitoring over several more years to firmly say anything about the activity cycle of any individual program star.

\subsection{Short-term Variability}\label{short}

When monitoring Ca II emission,
intrinsic stellar activity modulated by stellar rotation will appear along with any possible
activity stimulated by the planet. The orbital periods of the planets are well known and uniquely established by the PRV and transit discovery methods. The rotation periods of the stars are much harder to determine in part due to stellar differential rotation which yields non-unique periods. For our work, it is key to distinguish between the rotational and orbital modulation of chromospheric emission.

\begin{figure}
\epsscale{.70}
\plotone{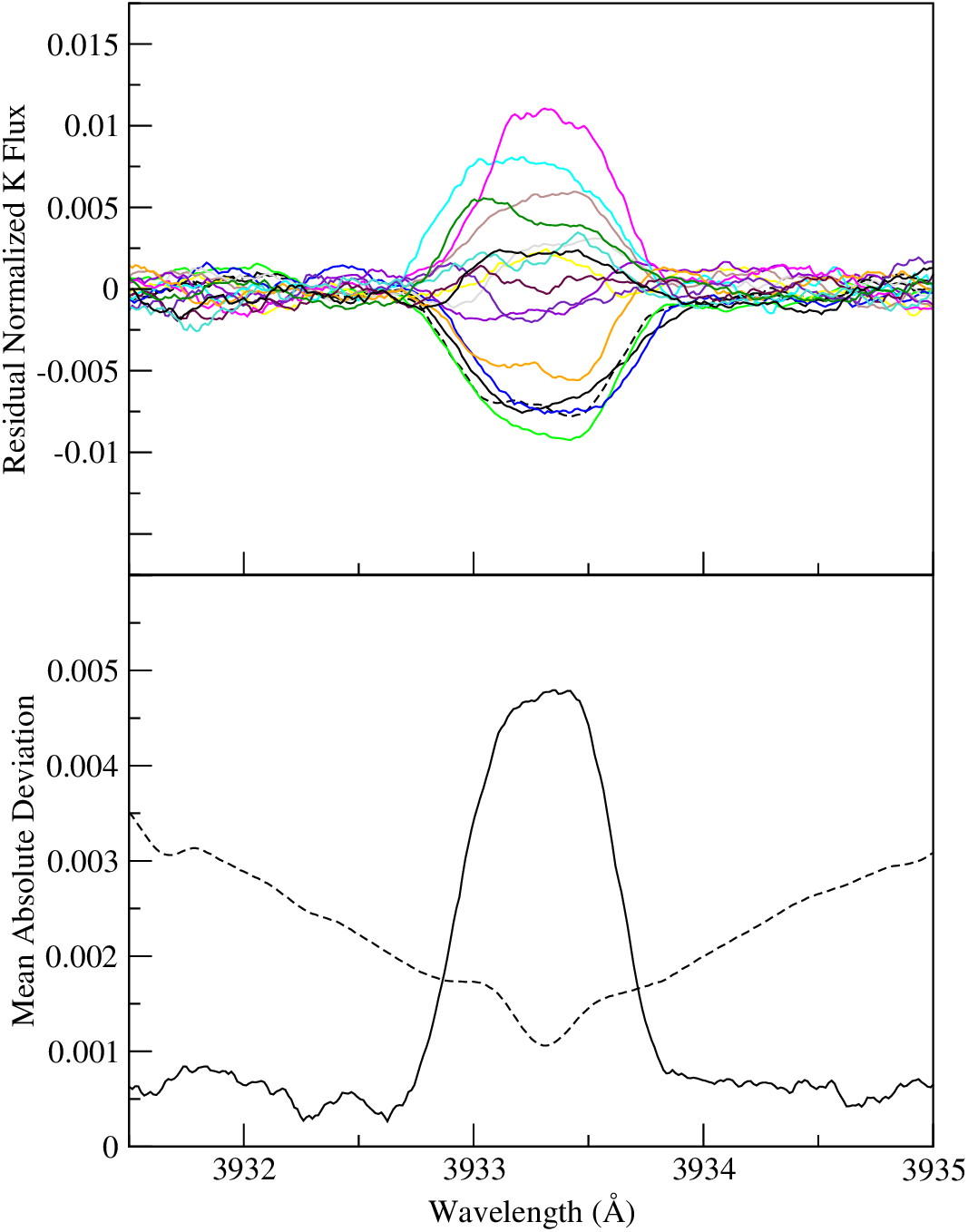}
\caption{Top: The residuals (smoothed by 21 pixels) from the normalized mean spectrum of the Ca~II~K core of $\upsilon$~And. Bottom: The mean absolute deviation (MAD) of the Ca II K core of $\upsilon$~And. The units are intensity as a fraction of the normalization level at 1/3 of the continuum. Overlaid (dashed line) is the mean spectrum (scaled down) indicating that the activity on $\upsilon$~And is confined to the K reversal.
\label{upsAnd_Kcombo}}
\end{figure}

To isolate the chromospheric activity within the reversals, we took nightly residuals from the average stellar spectrum of all data. Each residual spectrum had a broad, low-order curvature removed which was an order of magnitude less than the variations in the H and K lines discussed below. The residuals of the normalized spectra (smoothed by 21 pixels) were used
to compute the Mean Absolute Deviation (MAD = $N^{-1}\Sigma|data_{i}-mean|$ for $N$
spectra). As an example, the nightly residuals used to generate the MAD plot for $\upsilon$~And are displayed in Figure~\ref{upsAnd_Kcombo} (top). The MAD plot with the corresponding K-core superimposed is in Figure~\ref{upsAnd_Kcombo} (bottom). The
identical analysis described above was performed on the Ca~II H line of all target stars. The complete Ca II H results can be found in Shkolnik (2004). As
the two resonance lines share the same upper level and connect to the ground state, the same activity is seen in both. As expected, the Ca II H emission and activity levels are $\approx$~2/3 that of Ca II K (Sanford \& Wilson 1939).

\begin{figure}
\epsscale{1.0}
\plottwo{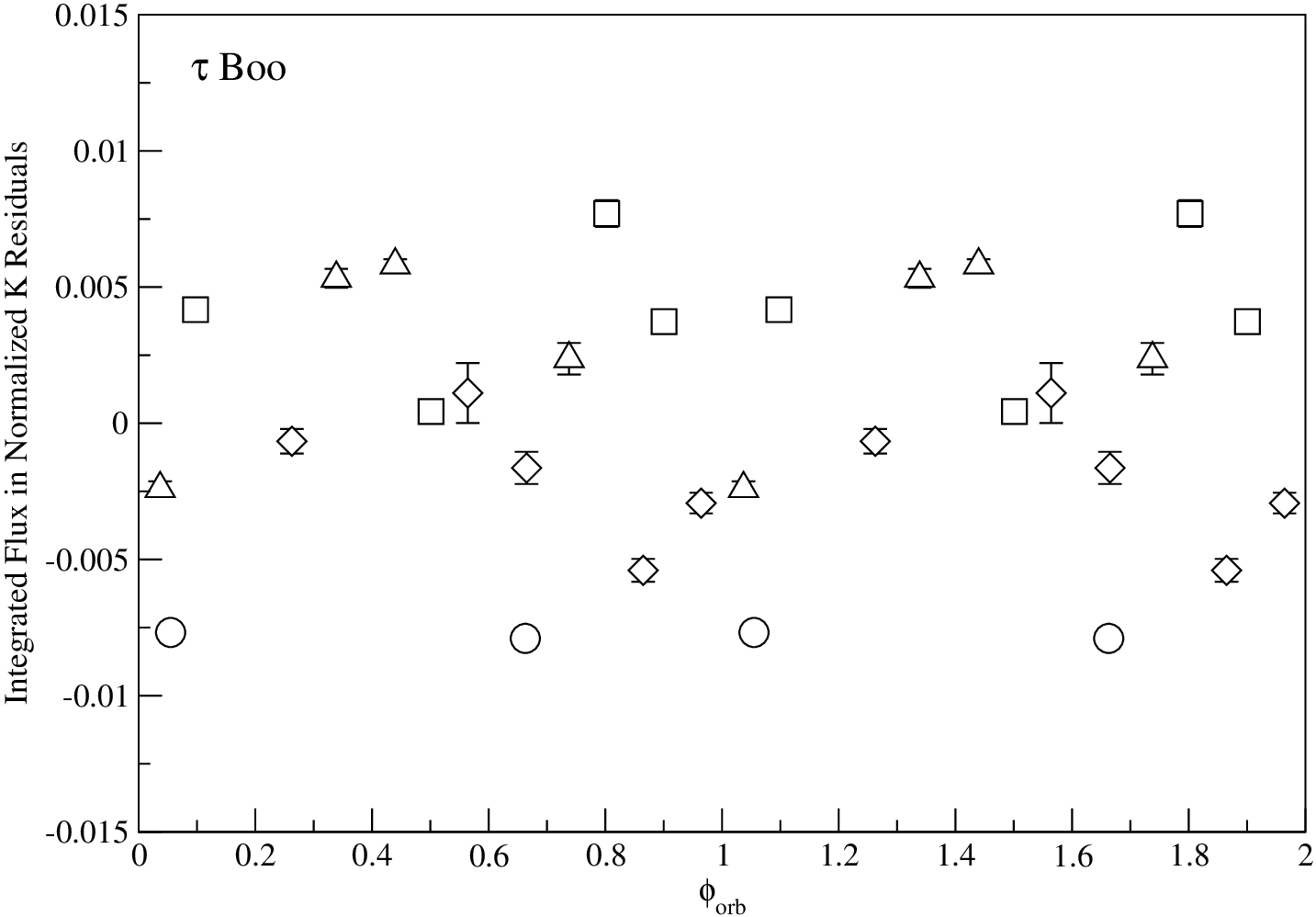}{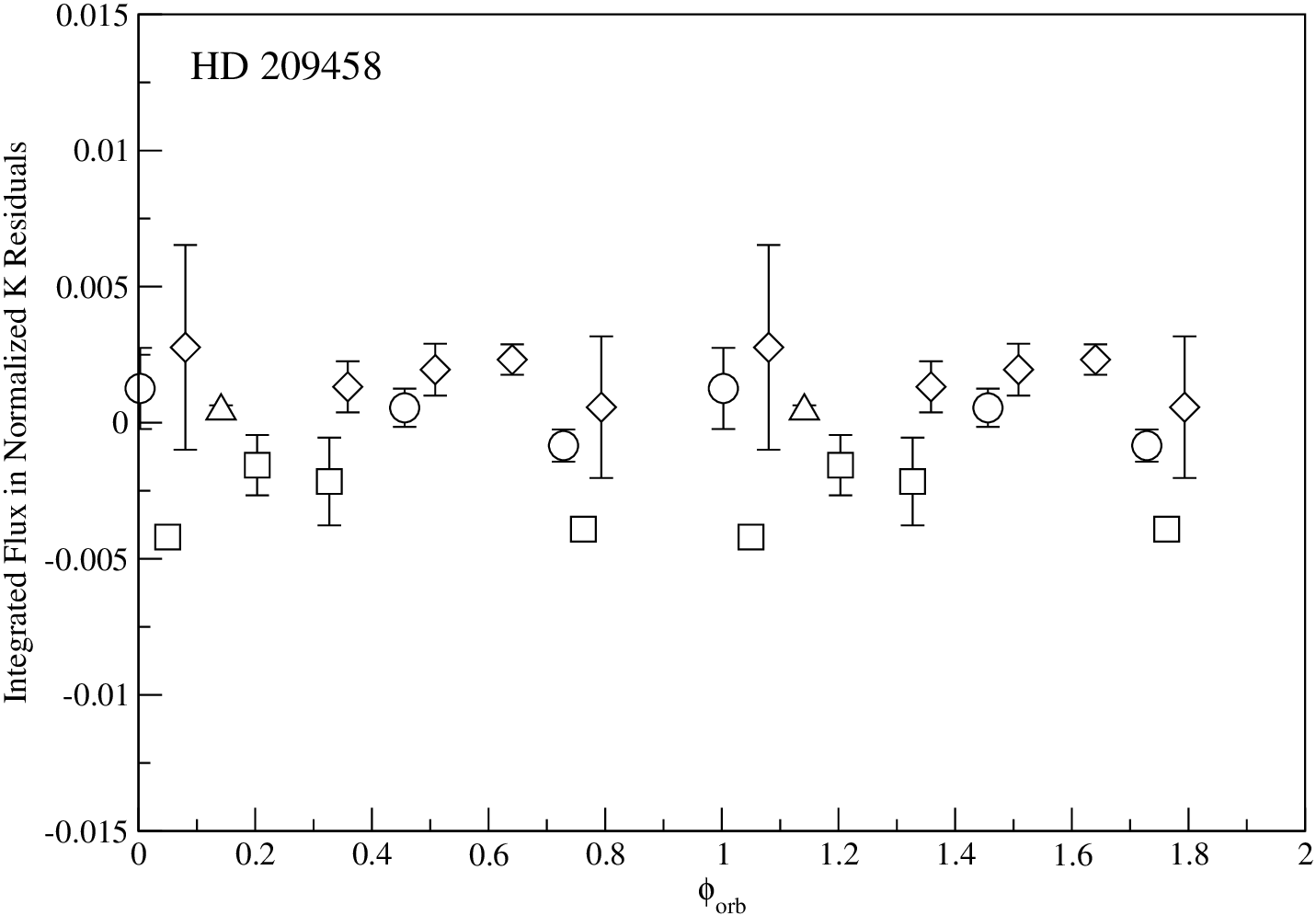}
\caption{Integrated flux of the K-line residuals from a normalized mean spectrum of $\tau$ Boo and HD~209458 as a function of orbital phase. The symbols distinguish data from different observing runs: circles $-$ 2001 August, squares $-$ 2002 July, triangles $-$ 2002 August, diamonds $-$ 2003 September.  Units of the integrated flux are in equivalent Angstroms
 relative to the normalization level which is approximately 1/3 of the stellar continuum. The error bars in residual flux are $\pm~1~\sigma$ as measured from the intra-night variations. The size of the phase error is within the size of the points.
\label{intK1}}
\end{figure}

The star with the shortest rotation period in our sample is
$\tau$~Boo.  It has the largest $v$sin$i$ (= 14.8 m~s$^{-1}$; Gray 1982) and is believed to be in
synchronous rotation with its tightly-orbiting planet ($P_{rot}$ = 3.2
$\pm$ 0.5 d, Henry et al.~2000; $P_{orb}$ = 3.31250 $\pm$ 0.00026 d, this work). If $\tau$~Boo is tidally locked to its planet, the planet-star interaction may be minimal (Saar et al.~2003, 2004) due to the fact that there is near zero relative velocity. (See Section~\ref{theory} for theoretical discussion.) The integrated residuals from the mean normalized K core are plotted against orbital phase in Figure~\ref{intK1} (left). Looking at the
individual observing runs in the plot for $\tau$~Boo, the two nights of
observation in 2001 August did not show much variation when the star
was somewhat less active. The 2002 July data
showed an increase in activity near the sub-planetary point relative to the
other observations in their respective runs.  However, in 2002 August and 2003 September observations show a relative
enhancement at $\phi\approx$ 0.4 $-$ 0.5 modulated with a period near $P_{rot}$ and $P_{orb}$.  This chaotic activity does
not allow us to draw any conclusions about planet-induced
heating.

Similar to $\tau$~Boo, the K emission of HD~209458 showed night-to-night modulation, but with
a smaller amplitude during most runs and without any phase coherence, as shown in Figure~\ref{intK1} (right).  In 2001 August, we caught the system immediately after
transit at which time we observed a slight enhancement in the Ca~II
emission relative to all other observations. In the 2002 July run, an
increase in emission occurred at $\phi\sim$ 0.25 with no apparent rise
toward $\phi$ = 0. Due to the
relatively low S/N of these data and the large intra-night deviations, we cannot form any conclusions.

As seen in Figures~\ref{intK1} and \ref{intK2}, four of the five CFHT stars show significant chromospheric variation throughout a single observing run. The standards, $\tau$~Ceti and the Sun, show no such modulation. One consistent result for the `active' stars ($\tau$~Boo, HD~179949, HD~209458 and $\upsilon$~And) is their night-to-night modulation of H \& K emission. Also, unlike the case for HD~73256 (see Section~\ref{modbyrot}), the night-to-night variations of these stars do not increase or decrease monotonically throughout an observing run, implying the variability cannot be explained exclusively with starspots rotating into or out of view. Another mechanism is necessary. The night-to-night variations may indicate planet-induced activity or sporadic flaring from hotspots.  If coupled to the planet, the localized activity would be travelling on the stellar surface faster than the star is rotating as it tracks the planet in its orbit.

Other than for $\tau$~Boo, the timescale
of activity is short compared to the stellar rotation
period.  Unfortunately, due to the large uncertainties in the rotation
periods of the other stars, phasing with rotation is
uninformative at this stage. However, we do know that $\upsilon$~And and HD~209458 both have $P_{rot}$ $>$~3$P_{orb}$ and rotate only $\lesssim 20^{\circ}$ per day. Wolf \& Harmanec (private communication) recently made UBV photometric
observations of HD 179949 at our request from the SAAO Sutherland
Observatory. When they combined their V observations with those from
Hipparcos (converted to V) they detected a rotation period of  7.07 d but
with an amplitude of only 0.008 mag. Given that the RMS of the V-observations was 0.006 mag, this periodicity is at the limit of detection.
Indirect indications of the rotation rate of HD 179949  imply $P_{rot} \approx 9$ days
and are presented in Shkolnik et al.~(2003). (See also Saar et al.~2004.) These include a high
X-ray luminosity for the star, a very long tidal synchronization timescale
and a moderate S$_{HK}$ index. While more photometry is needed to determine a
rotation period conclusively, it is highly unlikely that HD 179949 is
tidally locked to its planet at 3.092 d.

\subsubsection{Planet-induced Activity on Two Stars}\label{induced}

\begin{figure}
\epsscale{1.0}
\plottwo{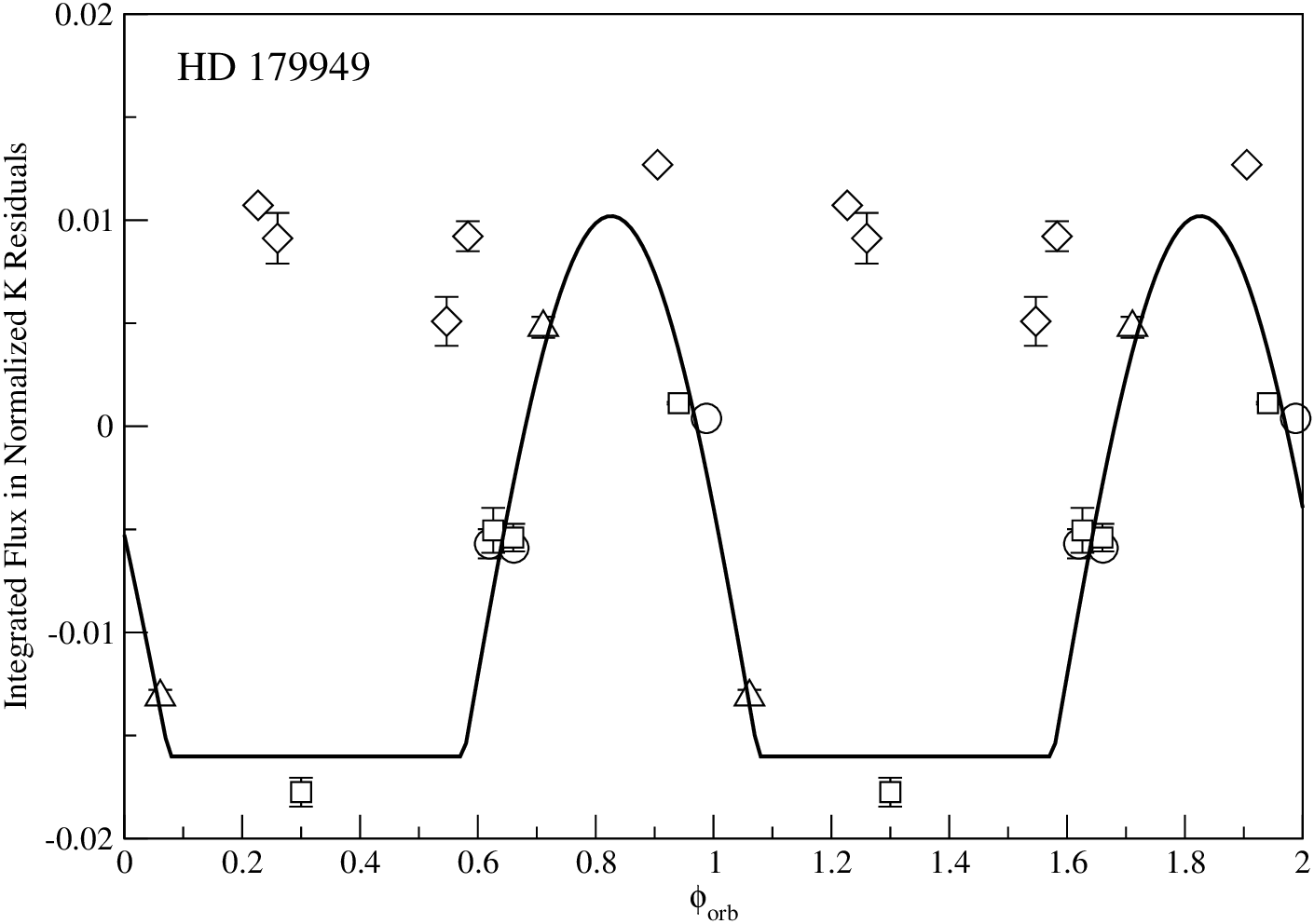}{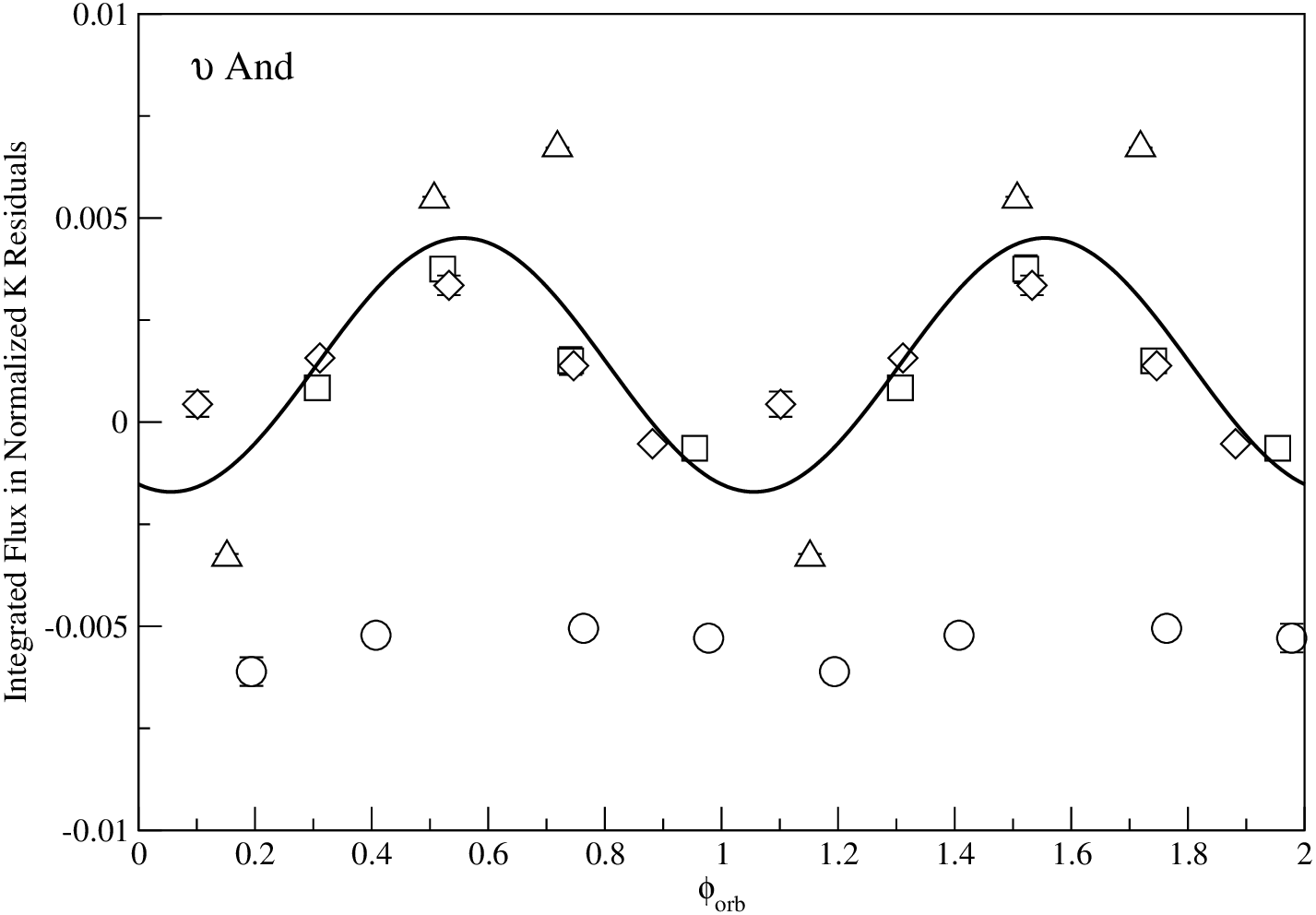}
\caption{Integrated flux of the K-line residuals from a normalized mean spectrum of HD~179949 and $\upsilon$~And as a function of orbital phase. The solid lines are best-fit spot models. For HD~179949, we used a truncated sine curve fitted to the 2001 and 2002 data and for $\upsilon$ And, we fitted the curve to the 2002 and 2003 data.
\label{intK2}}
\end{figure}

In Shkolnik et al.~(2003) we presented the first evidence of planet-induced heating on HD~179949. The effect lasted for over a year and peaked only once per orbit, suggesting a magnetic interaction. We fitted a truncated, best-fit sine curve with $P = P_{orb}$ =
3.092 d corresponding to the change in projected area of a bright spot on
the stellar surface before being occulted by the stellar limb. Figure~\ref{intK2} (left) updates the
integrated K residuals to include the 2003 September data.  The spot model is a remarkable fit for the 2001 and 2002 data peaking at $\phi$ = 0.83 with an amplitude of 0.027.
Clearly, the average K emission is higher during the latest run (as shown in Figure~\ref{yearly}) with a much smaller level of variability. It is interesting to note that the 2003 data still peak between $\phi$ = 0.80 $-$ 0.95, consistent with the previous results.

The second convincing case of magnetic interaction is between $\upsilon$~And and its innermost giant planet.\footnote{$\upsilon$~And has three known Jupiter-mass planets at 0.059, 0.829, and 2.53 AU (Butler et al.~1999).} In Figure~\ref{intK2} (right), the 2002 July, 2002 August and 2003 September runs show good agreement in
phase-dependent activity with an enhancement at $\phi$ = 0.53. The best-fit sine curve has an amplitude of 0.0032. The 2001 August fluxes are lower than the mean of all four observing
runs by almost 3\% and still display a significant ($>$~2$\sigma$) modulation like the quiescent epoch of HD~179949. Again, even the low-amplitude modulation has a rise and fall with a period consistent with $P_{orb}$ and peaks near $\phi~=0.5$.

For these two cases, the peak of the emission does not directly coincide with the sub-planetary point, $\phi$ = 0.  For HD~179949, it leads the planet by 60$^{\circ}$ in
phase and for $\upsilon$~And, the Ca II emission is 169$^{\circ}$ out of phase with the sub-planetary point. Santos et al.~(2003) also observed a 90$^{\circ}$ lag from the sub-binary point in the periodic activity indicated by the photometric variations for HD 192263. The phase lead or lag may help identify the nature of the interaction.  For example, the phase offset of a starspot or group of starspots can be a characteristic effect of tidal friction, magnetic drag or reconnection with off-center stellar magnetic field lines, including a Parker spiral-type scenario (Weber \& Davis 1967, Saar et al.~2004). In any case, the phasing, amplitude and period of the activity have persisted for over a year between observations.  For HD~179949, this equals 108 orbits or at least 37 stellar rotations and for $\upsilon$~And, the time spans 88 orbits or approximately 29 rotations.

The observations are consistent with a magnetic heating scenario as the chromospheric enhancement occurred only once per orbit. We estimated the excess absolute flux released in the enhanced chromospheric emission of HD~179949 by calibrating the flux with that of the Sun. The flux was the same order of magnitude as a typical solar flare, $\sim$~10$^{27}$ erg~s$^{-1}$ or 1.5$\times$10$^{5}$ ergs~cm$^{-2}$~s$^{-1}$. This implies that flare-like activity triggered by the interaction of a star with its hot jupiter may be an important energy source in the stellar outer atmosphere. This also offers a mechanism for short-term chromospheric activity on the stars with close-in Jupiter-mass planets.

\subsubsection{The Non-Varying Program Stars}\label{non_varying}

\begin{figure}
\epsscale{1.0}
\plottwo{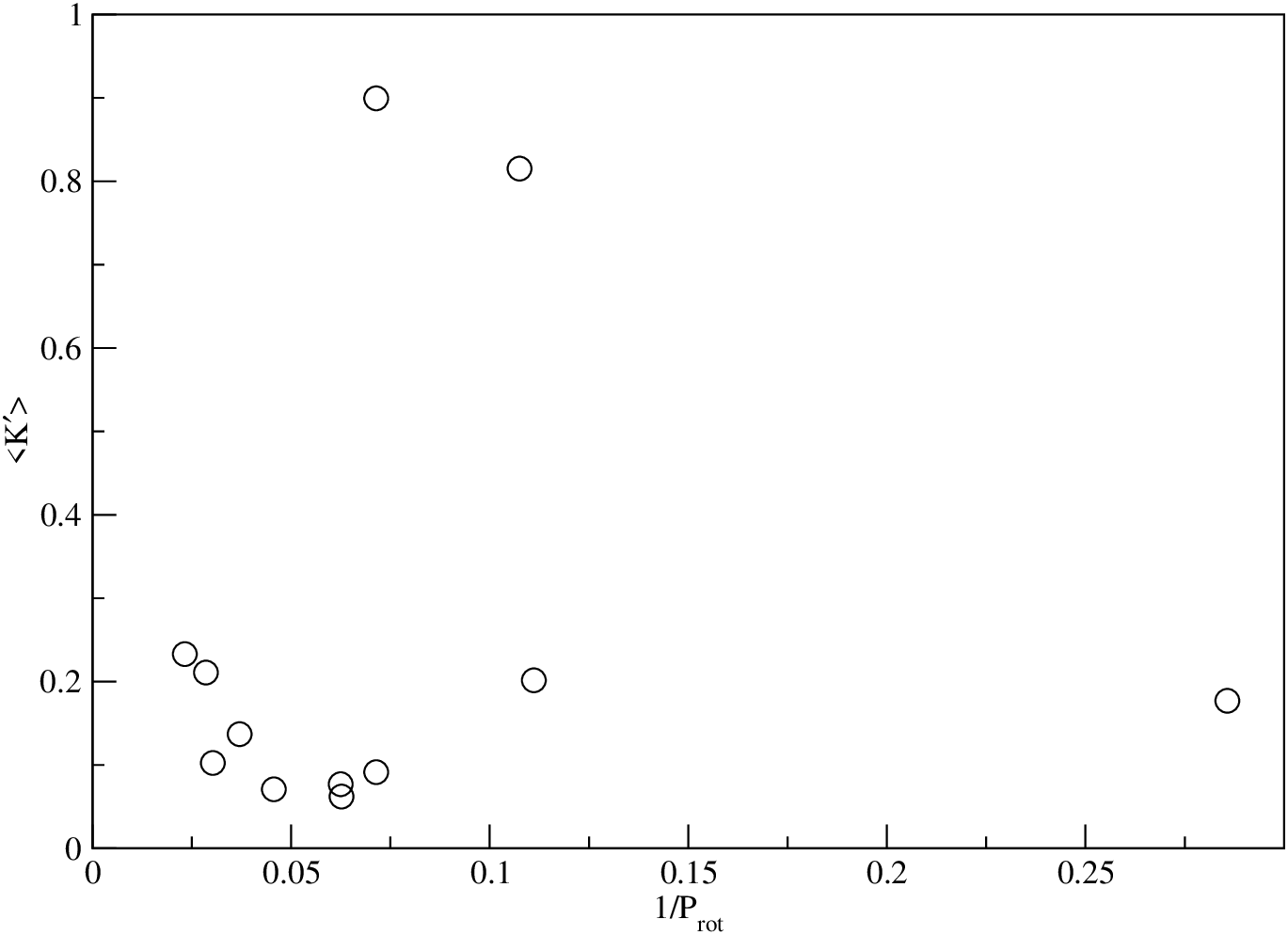}{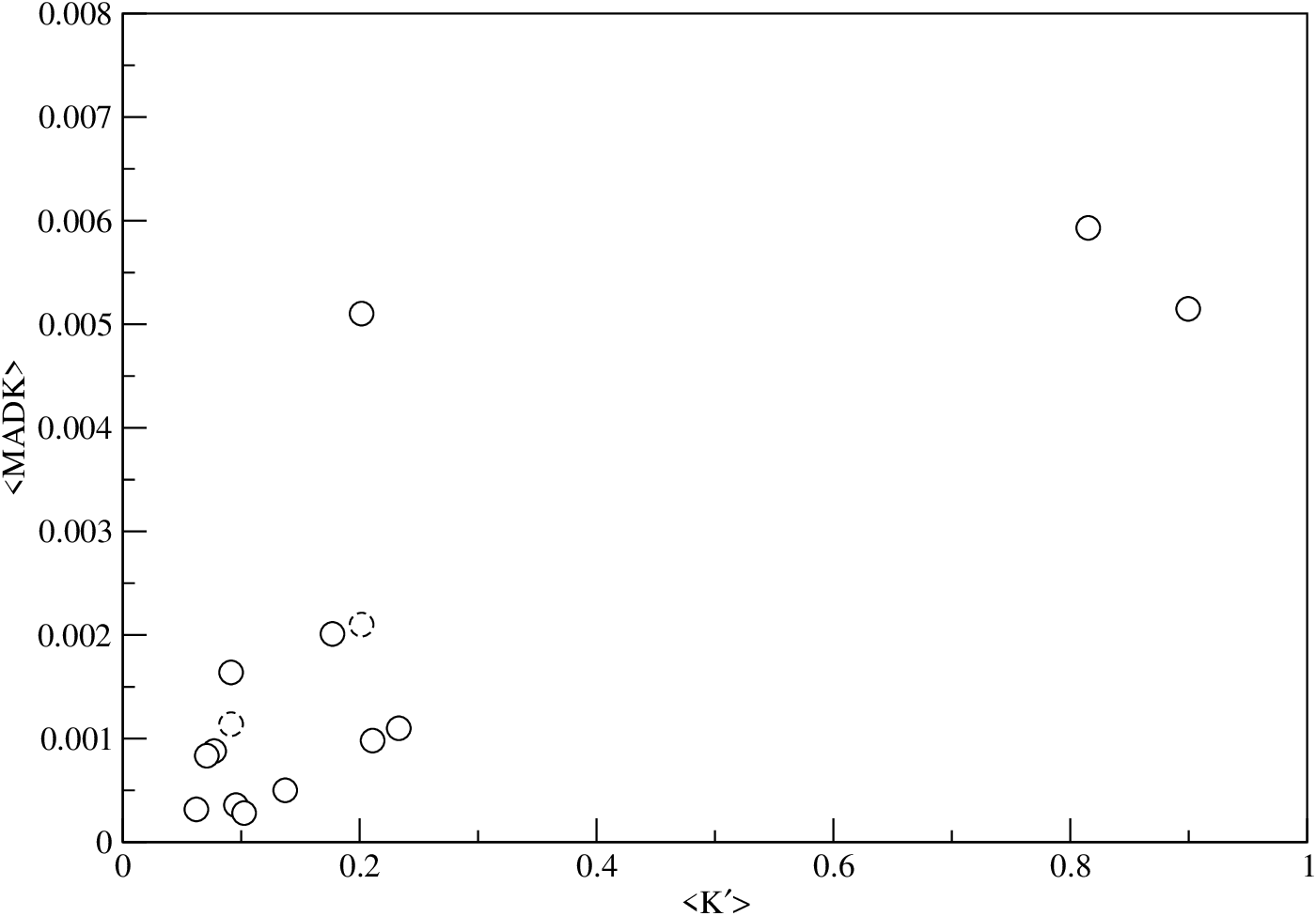}
\caption{Left: 1/$P_{rot}$ in days plotted against $\langle$K$\arcmin\rangle$, the mean chromospheric emission in the K line, for all stars less HD 76700 for which no $P_{rot}$ is published. Right: $\langle$K$\arcmin\rangle$ as a function of the mean MAD K values per run for all the stars. The dashed circles are $\langle$MADK$\rangle$ for HD~179949 and $\upsilon$~And with the orbital (geometric) modulation removed. Units for $\langle$K$\arcmin\rangle$ and $\langle$MADK$\rangle$ are in equivalent angstroms relative to the normalization level.
All values are listed in Table~\ref{starpars}.
\label{MADK}}
\end{figure}

Of the 10 program stars we monitored for H \& K variability, five of them showed no changes down to the 0.001 level: 51~Peg, HD~46375, HD~75289, HD~76700, HD~83443. There are two reasons we offer to explain the relative quiescence of these stars.  It was well-known through the many years of the Mt.~Wilson S$_{HK}$ survey there is a strong correlation between rotation rate (or inversely with rotation period) and Ca II emission (Noyes et al.~1984, Pasquini et al.~2000).  This is a likely contributor though is not obviously clear in our sample set as shown in Figure~\ref{MADK} (left) where the inverse of the rotation period is plotted against the mean chromospheric emission of the Ca II K line $\langle$K$\arcmin\rangle$.
The photospheric contribution to the emission was removed from $\langle$K$\rangle$ using $\langle$K$\arcmin\rangle$ = $\langle$K$\rangle$(1 - ${R_{phot} \over R_{HK}}$) where R$_{phot}$ = R$_{HK}$ $-$ R$\arcmin_{HK}$ and is an empirical function of ($B - V$) and S$_{HK}$ taken from Hartmann et al.~(1984). The chromospheric contibutions R$\arcmin_{HK}$ are from Wright et al.~(2004).\footnote{For those few stars that were not in Wright et al.'s paper, we removed the photospheric contribution as tabulated from stars in Wright et al.'s sample of the same spectral type and log($g$).}  In Table~\ref{starpars} we list $\langle$K$\rangle$, $\langle$K$\arcmin\rangle$, and $\langle$MADK$\rangle$ for all the stars. From Figure~\ref{MADK} (right), we deduce that the higher a star's chromospheric emission, the more night-to-night activity it displays. Radick et al.~(1998) show the same effect for a much larger sample. This is akin to shot noise since the flaring or stochastic noise associated with the activity will increase with the activity level.

\begin{figure}
\epsscale{0.65}
\plotone{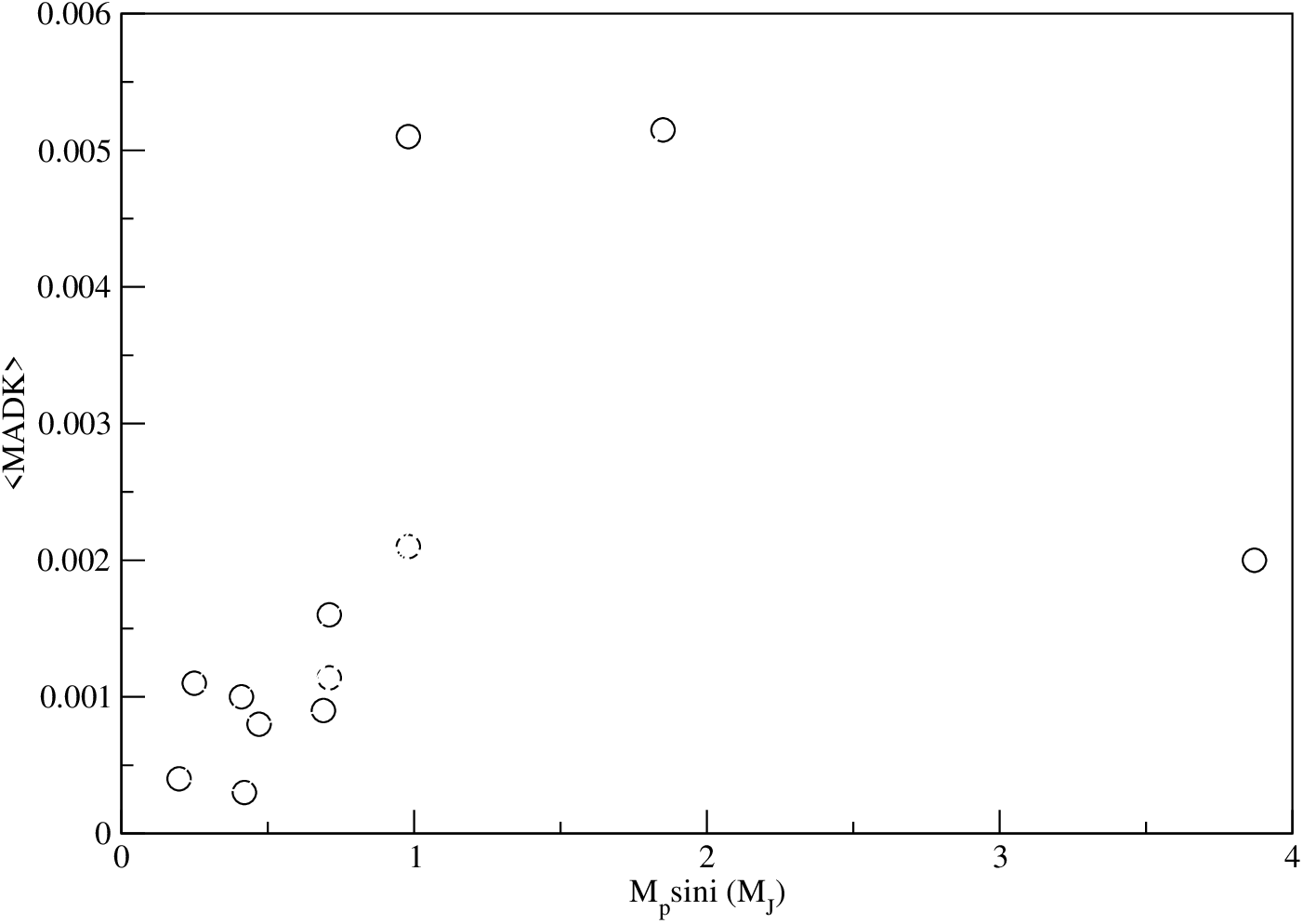}
\caption{The minimum planetary mass (in Jupiter masses) plotted against the average MAD of the K-line per observing run. The dashed circles are $\langle$MADK$\rangle$ for HD~179949 and $\upsilon$~And with the orbital (geometric) modulation removed.
All data are listed in Table~\ref{starpars}.
\label{msini_MADK}}
\end{figure}

Secondly, a recent calculation of the magnetic fields in giant extrasolar planets (S\'anchez-Lavega 2004) looked at the internal structure and the convective motions of these planets in order to calculate the dynamo-generated surface magnetism. Given the same angular frequency (which is a reasonable approximation for the short-period planets in question), the magnetic dipole moment, and hence the magnetospheric strength, increases with planetary mass. This is observed in our own solar system for the magnetized planets where the magnetic moment grows proportionally with the mass of the planet (Arge et al.~1995).  Since only lower limits exist for most of the hot jupiters, we can only plot $M_{p}$sin$i$ against $\langle$MADK$\rangle$ in Figure~\ref{msini_MADK} where we still see an intriguing correlation.
The dashed circles for HD 179949 and $\upsilon$ And are their $\langle$MADK$\rangle$ values (0.0021, 0.0011, respectively) with the orbital modulation removed.  Of our sample, $\tau$ Boo has the most massive planet and yet falls well below the correlation. As we discuss further in Section~\ref{theory}, if the star and planet are tidally locked, as is thought to be the case for $\tau$ Boo, then there is little or no free energy left from the orbit and we would expect weak, if any, magnetic coupling.

\subsection{Modulation by Rotation}\label{modbyrot}

For most of our target stars, rotation periods are not well enough known to accurately phase the Ca II data with $P_{rot}$.  There is even ambiguity in $P_{rot}$ of the often-observed $\kappa^{1}$~Ceti. In late 2003, $\kappa^{1}$~Ceti was monitored continuously for 30.5 days by MOST.\footnote{The MOST microsatellite (Microvariability \& Oscillations of STars), a Canadian photometric telescope recently launched to observe p-mode oscillations on sun-like stars (Walker et al.~2003b)}  This best-ever lightcurve obtained for $\kappa^{1}$~Ceti showed a pattern composed of two transiting spots of differing periods, 8.9 and 9.3 days providing a direct and unique measurement of differential rotation (Rucinski et al.~2004). The real ambiguity in $P_{rot}$ from the space observations of $\kappa^{1}$ Ceti demonstrates the difficulty in measuring $P_{rot}$ for most stars.

Nonetheless, two stars do exhibit clear rotational modulation: $\kappa^{1}$~Ceti, a star with no confirmed planet (Halbwachs et al.~2003) and HD~73256, a star with a 1.85-M$_{J}$ planet orbiting at 0.037~AU (Udry et al.~2003).  The Ca II emission from $\kappa^{1}$~Ceti has been monitored by Baliunas et al.~(1995) through the narrow-band filter of the Mt.~Wilson survey from 1967 to 1991.  These data show long-term stability of a period of 9.4 $\pm$ 0.1 d, close to the photometric rotation period of 9.214 d published by Messina \& Guinan (2002). The rotation period for HD~73256 is photometrically determined to be 14 d (Udry et al.~2003), consistent with the 13.9 days derived from the R$\arcmin_{HK}$ activity index (Donahue 1993).

\begin{figure}
\epsscale{1.0}
\plottwo{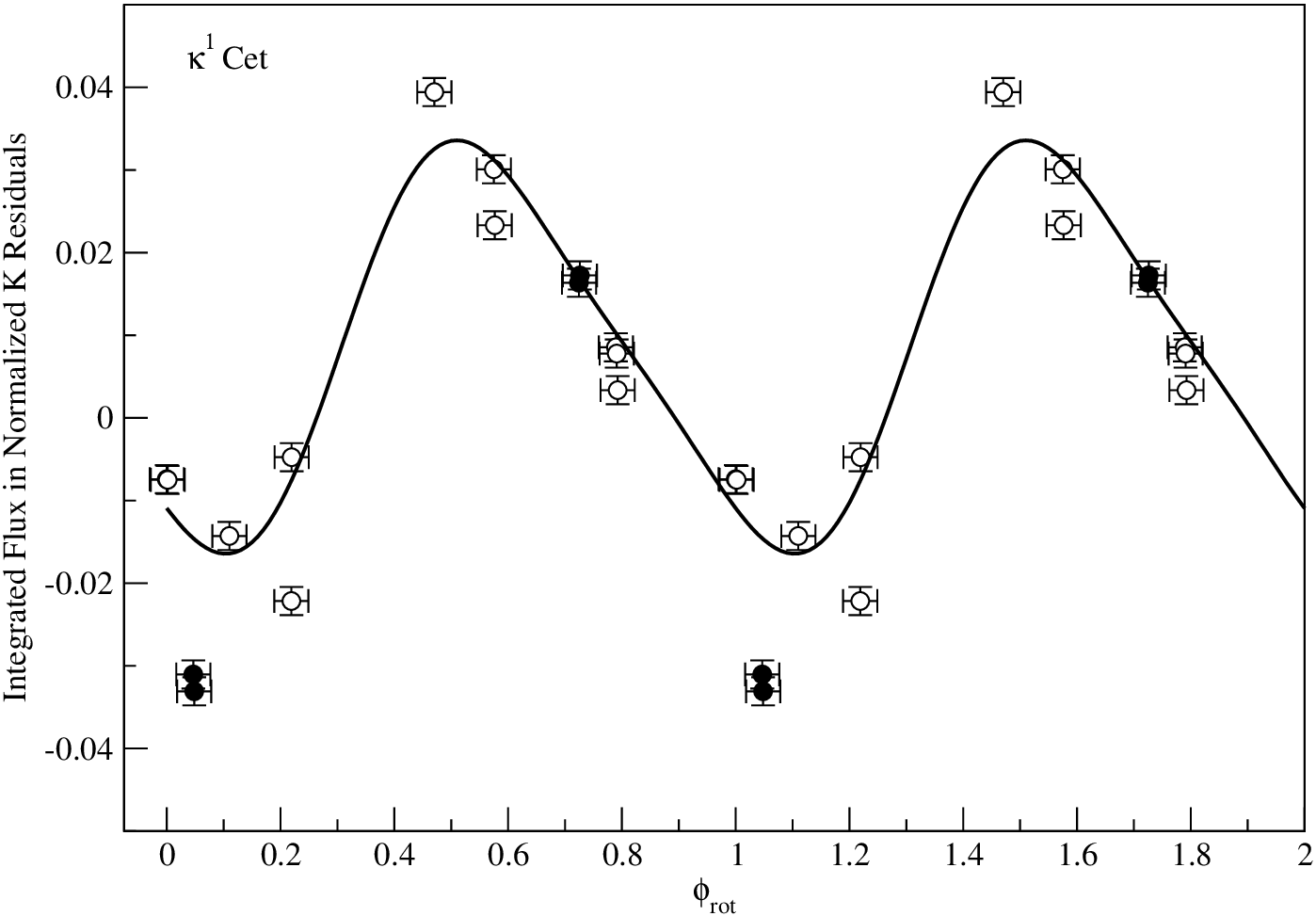}{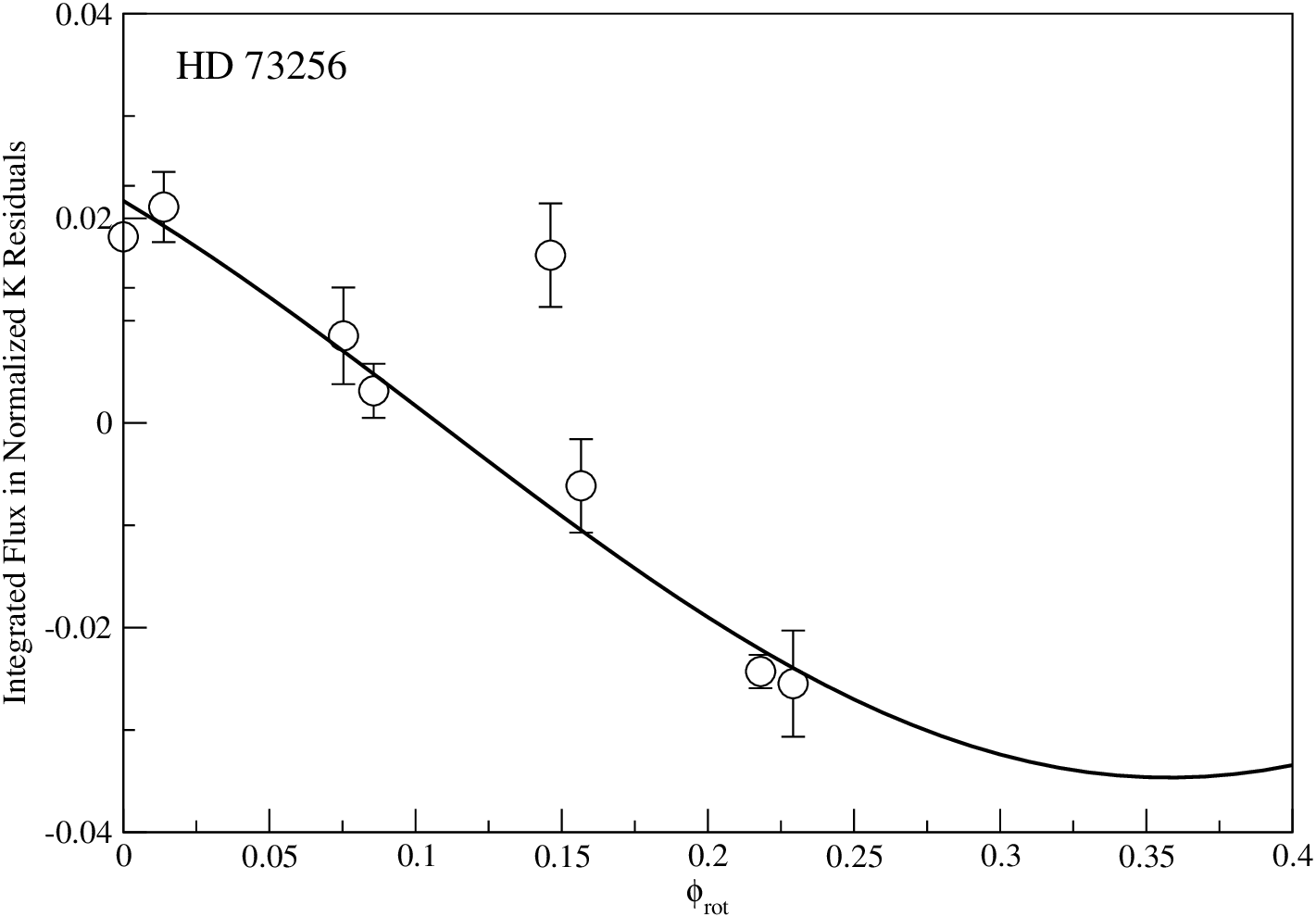}
\caption{Integrated flux of the K-line residuals from a normalized mean spectrum of HD~73256 and $\kappa^{1}$~Ceti as a function of rotational phase.  The solid lines are the best-fitting periodical curves.  In the $\kappa^{1}$~Ceti plot, the open circles are 2002 data and the closed circles are 2003 data.
\label{intK_Prot}}
\end{figure}

\begin{figure}
\epsscale{0.7}
\plotone{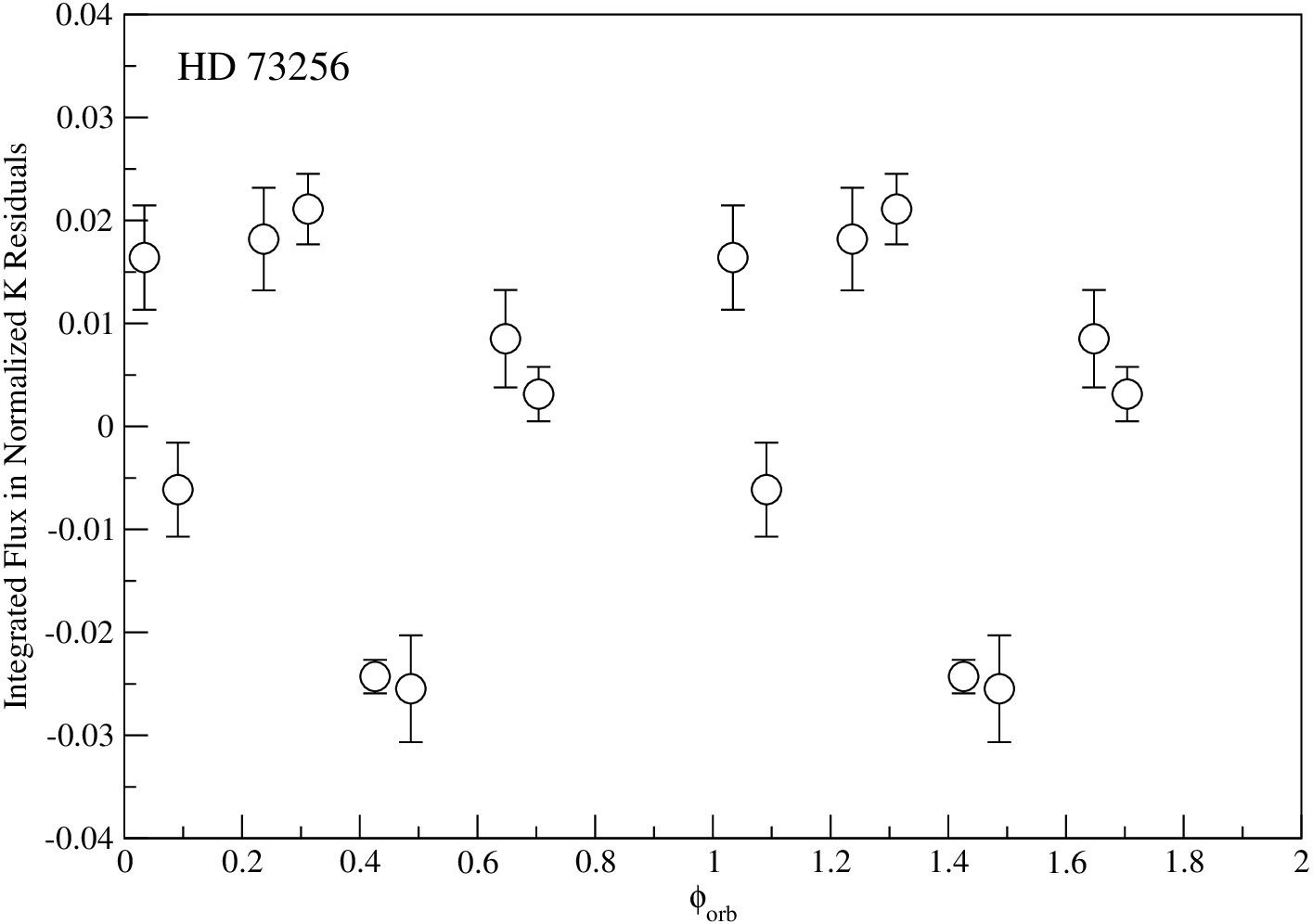}
\caption{Integrated flux of the K-line residuals from a normalized mean spectrum of HD~73256 as a function of orbital phase.
\label{hd73256_intK}}
\end{figure}

We observed periodic Ca II H \& K variability in the chromosphere of $\kappa^{1}$~Ceti during our 2002 and 2003 CFHT runs from which we determined a rotation period of 9.332 $\pm$ 0.035 d. The results were first published in Rucinski et al.~(2004) where we compared the activity seen in Ca II with MOST's lightcurve showing that the chromospheric activity coincided with the low-latitude spot such that the maxima of the two curves agree. HD~73256 was observed twice per night for four nights at the VLT. The mean K cores of these two stars are shown at the top of Figure~\ref{vlt_Kcores} where their similarly strong emission is evident.  The high level of chromospheric emission points to a young age for both of these stars:  650 $-$ 750 Myr for $\kappa^{1}$~Ceti (G\"udel et al.~1997; Dorren \& Guinan 1994) and 830 Myr for HD~73256 (Donahue 1993).  The $\langle$MADK$\rangle$ given for these two stars in Table~\ref{starpars} are corrected for the rotational modulation resulting in activity levels comparable to HD~179949. The integrated residual K fluxes for $\kappa^{1}$~Ceti and HD~73256 are plotted against relative rotational phase in Figure~\ref{intK_Prot}. For completeness, we plot the residual K fluxes of HD~73256 as a function of orbital phase in Figure~\ref{hd73256_intK} where the flare is apparent at $\phi_{orb}$ = 0.03. There is no clear signature of the planet in the activity.

In both cases, $\kappa^{1}$~Ceti ($M_{V}$ = 4.92) and HD~73256 ($M_{V}$ = 5.27) show sporadic flaring beyond the clear rotational modulation. The periodic best-fit curve for $\kappa^{1}$~Ceti necessitates a non-zero eccentricity while a sine curve is sufficient for HD 73256. The largest excursion from the rotation curve of $\kappa^{1}$~Ceti has an energy of 2.8$\times$10$^{4}$ erg~cm$^{-2}$~s$^{-1}$ (again, measured by comparing with solar absolute flux). We estimate the absolute flux emitted from HD~73256's flare at $\phi_{rot}$ = 0.15 to be 4.9$\times$10$^{4}$ erg~cm$^{-2}$~s$^{-1}$ (or $>2.9\times10^{8}$ erg~cm$^{-2}$ if the flare lasted for at least the hour for which we observed it.)  The modulation of the K emission due to rotation is $\approx$~6\% indicating that the emission is dominated by a large hotspot on the stellar surface.  As we have seen on HD~179949 and $\upsilon$~And, planet-induced variations are at the level of 1 $-$ 2\% suggesting that the reigning hotspot could have diluted any heating caused by the hot jupiter.

\begin{figure}
\epsscale{0.65}
\plotone{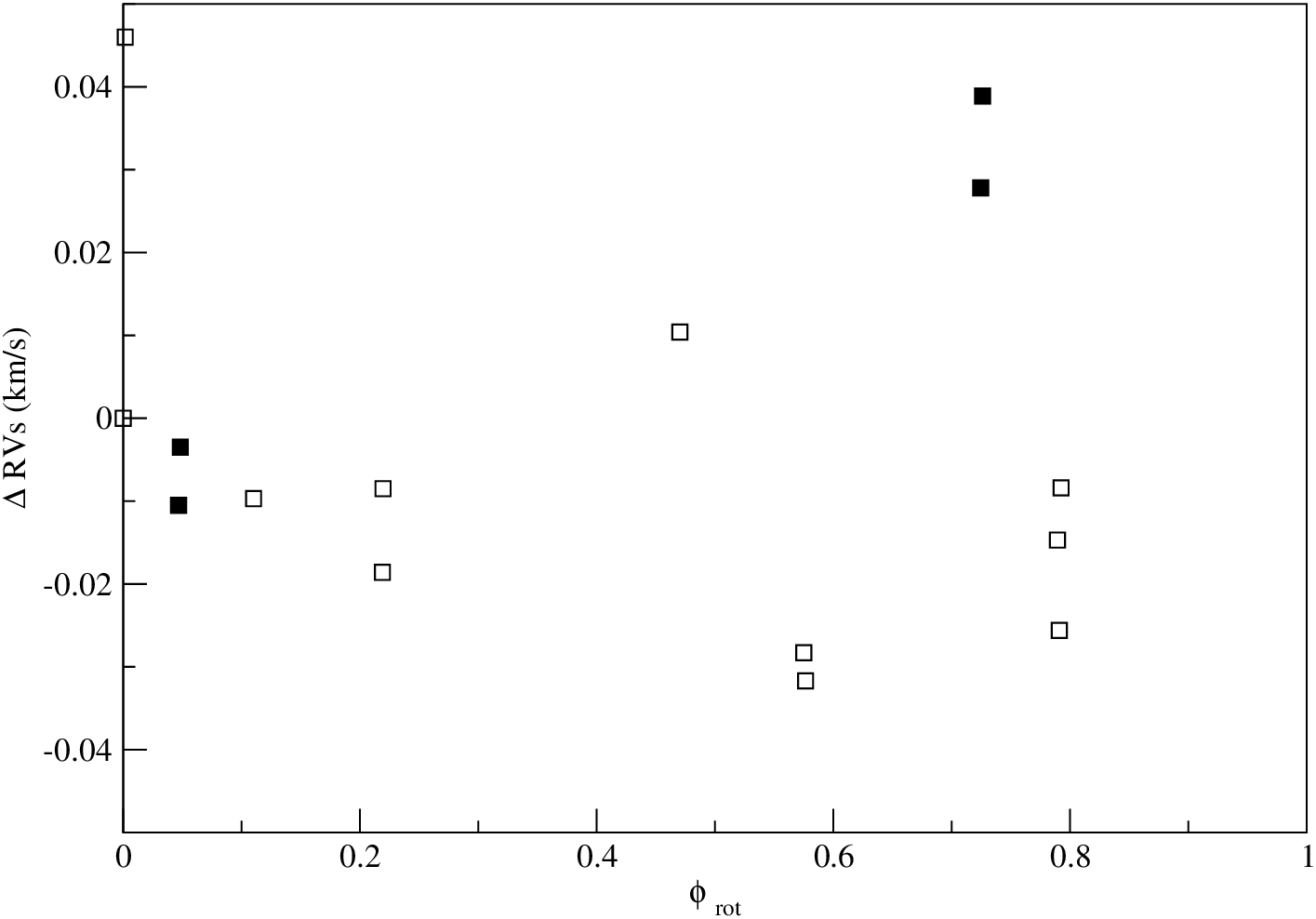}
\caption{Relative radial velocities for $\kappa^{1}$~Ceti.  Open squares are the 2002 data and the filled in squares are the 2003 data.
\label{rv_k1ceti}}
\end{figure}

The $\Delta$RVs for $\kappa^1$~Ceti are plotted in Figure~\ref{rv_k1ceti} against the 9.332-d phase determined from the K-line residuals  where the open symbols correspond to data from 2002 and the solid squares to 2003. From the 2002 data we derive $\sigma_{RV} = $21.8 m~s$^{-1}$ and 23.6 m s$^{-1}$ when combined with 2003. These values are very similar to
$\sigma_{RV} = $24.4 m s$^{-1}$, found over 11 years by Cumming et al.~(1999). In 2003 the $\Delta$RVs appear significantly different between the two nights, something which seems to be reinforced by the consistency within the pairs of $\Delta$RVs. While a planetary perturbation cannot be ruled out by the 2002 data and other PRV studies, the difference in 2003 might be associated with the velocity field of the star itself. The increase of velocity with increasing K-line strength at $\phi_{rot} \approx$ 0.7 is consistent with the extreme event in 1988 seen by Walker et al.~(1995).
However it should be emphasized, the possibility remains of a close giant planet around $\kappa^1$~Ceti. For instance, a planet inducing a reflex radial velocity variation $< 50$~m~s$^{-1}$ and
tidally synchronized with the star, would have $M_{p} \sin i \simeq 0.74$~M$_J$
and $a$ = 0.084 AU.

\section{A Physical Scenario}\label{theory}
The enhancements of chromospheric activity on HD 179949 and $\upsilon$ And appear only once per orbit implying a magnetic, rather than tidal, interaction between the star and its hot jupiter.
The two stars are F-type stars
with higher X-ray luminosity than the solar value. The ROSAT catalogue of bright main-sequence stars lists HD~179949 as having at least double the X-ray luminosity (a measurement independent of $i$) of most other single F8 $-$ 9 dwarfs (H\"unsch et al.~1998). Yohkoh solar X-ray observations (e.g. Yokoyama \& Shibata 2001) have
shown that the energy release at the site of magnetic reconnection
during a solar flare generates a burst of X-ray emitting gas.  This
hot plasma is funneled along the magnetic field lines down to the
surface producing `footprints' through anomalous heating of the Sun's
chromosphere and transition region.  It is this same phenomenon that
is likely occurring between hot jupiters and their host stars through the
reconnection of their magnetic fields.  Observationally, companion-induced activity is unambiguously observed on RS CVn stars, as discussed in Section~\ref{intro}.

The hot jupiter of
$\upsilon$ And is located farther out from the host star
than that of HD 179949, implying that the
magnetic interaction between the star and its hot jupiter
is diminished. This would result in less Ca~II enhancement as is shown in our data. $\tau$~Boo is
also an F-type star with intense X-ray emission, but its magnetic
influence is likely reduced by the small relative motion in the azimuthal direction
between the planet and the stellar magnetosphere in an almost
final equilibrium state in which the star and the planet
are tidally locked to each other. While more data are required
to truly verify whether the variability of Ca~II emission is
correlated to the apparent position of the hot jupiter in these systems,
we explore and review a few theoretical aspects of planet-induced heating
scenarios in this section.

The location of the hot jupiter relative to the Alfv\'en radius
(the distance from the star at which the radial velocity of
the wind $V_{r,{\rm wind}}$ equals the local Alfv\'en velocity
$V_A$) plays a significant role in transporting energy toward the
star against the stellar winds.
Since the Alfv\'en radius
of the Sun is about 10 to 20 R$_{\odot}$ at solar minimum and
30 R$_{\odot}$ at solar maximum (e.g.~Lotova et al.~1985),
the small distance $\lesssim 0.1$ AU of hot jupiters
from their host stars suggests that unlike our Jupiter,
surrounded by a bow shock, some of these
hot jupiters are located inside the Alfv\'en radius depending
on the magnetic strength of their host stars (Zarka et al.~2001,
Ip et al.~2004). Therefore the direct magnetic interactions
between a hot jupiter and its star without a bow shock
might resemble the Io-Jupiter interactions (Zarka et al.~2001)
or the RS CVn binaries (Rubenstein \& Schaefer 2000).
Most of the theoretical models applied to these two cases have focused on the
geometry of intertwined magnetic fields as well as the
the energy transport through Alfv\'en waves and/or induced
currents.

Alfv\'en waves cannot propagate along the stellar
field lines toward the star
in the region outside the Alfv\'en radius
where the group velocity of
Alfv\'en waves is always in the positive radial
direction (e.g. Weber \& Davis 1967).\footnote{Since the sound
speed is much less than the Alfv\'en speed in the extended coronal
region, the fast Alfv\'en radius is almost the same as the Alfv\'en radius.
Therefore we do not distinguish these two radii in this paper.
However unlike the Alfv\'en modes channeling along field lines, the fast Alf\'ven waves
can propagate in all directions inside the Alf\'ven radius
and therefore their impact on the parent star is attenuated.
Here we assume that the Alfv\'en mode is the dominant, or
at least the comparable, mode compared to the other compressible
modes when they are excited.} Other means of inward energy transport
require their energy flux to be at least larger than the energy
flux carried by the stellar winds.

Ip et al.~(2004) estimate the input magnetic power due to
the relative motion between the synchronized hot jupiter and
the stellar magnetosphere
to be $\sim 10^{27}$ ergs~s$^{-1}$,
the same order of magnitude of a typical solar flare. A
similar amount of power might be obtained
based on the induced current model (Zarka et al. 2001)
if the radius of the
ionosphere of an unmagnetized hot jupiter in the induced current model
can be approximated by the radius of the magnetopause of a magnetized
hot jupiter. The observed excess energy flux from an unresolved
disk of the star is equal to this input magnetic power
averaged over the disk of the star. That is, the energy
flux is roughly equal to
\begin{eqnarray}
&&(B_m^2/8\pi)
(V_{orb}-V_{\phi,{\rm wind}}) (r_m/R_*)^2 \nonumber \\
&=&
\left( {B_*^{2(1-1/q)} B_p^{2/q} \over 8\pi} \right)
a \left( {2\pi \over P_{orb}}-{2\pi \over P_{rot}} \right)
\left( {R_* \over a} \right)^{2p(1-1/q)}
\left( {R_p \over R_*} \right)^2,
\label{eq:energyflux}
\end{eqnarray}
where $r_m$ is the radius of the planet's magnetopause,
$B_{m}$ is the magnetic field at the magnetopause,
$R_*$ is the radius of the star, $R_p$ is the radius of
the planet, $a$ is the distance between the star and the
planet,
$V_{orb}$ and $P_{orb}$ are the orbital
velocity and period respectively, $P_{rot}$ is the rotation period
of the star, $V_{\phi,{\rm wind}}$ is
the azimuthal component of the stellar wind
velocity, and $B_*$ and $B_p$ are the mean magnetic field
on the surface of the star and the planet. In deriving the above equation, we have assumed that the
stellar magnetic field decays as $r^{-p}$, the planetary field
decays as $r^{-q}$, $r_m$ was determined by equating the
stellar and planetary fields at the magnetopause (i.e.
$B_*(R_*/a)^p = B_p(R_p/r_m)^q$), and the stellar magnetosphere
inside the Alfv\'en radius is nearly co-rotating with the
stellar rotation (i.e. $V_{\phi,{\rm wind}}\approx a\Omega_*$).
The radial component of the stellar wind $V_{r,{\rm wind}}$ is left
out from the above estimate as long as the condition
$V_{r,{\rm wind}} \lesssim V_{orb}-V_{\phi,{\rm wind}}$
is valid.
The energy flux in eq(\ref{eq:energyflux})
should be regarded as the maximal input energy from the
planet's orbital energy because only some fraction of this
amount of energy is transferred to the Ca II emissions.
For $p=2$ (Vrsnak et al.~2002 for the
case of our Sun, and Weber \& Davis 1967 for the case
of open fields), $q=2$,
$a$ = 0.045~AU, $R_*=1.3$~R$_{\odot}$, $R_p=1.1$~R$_J$,
$P_{rot}=9$ d, $B_{*}$ = 200~G, and $B_{p}$ = 10~G,
the energy flux given by eq(\ref{eq:energyflux}) is
roughly equal to
10$^{5}$ erg~cm$^{-2}$~s$^{-1}$,
a value comparable to the differential intensity
from our data of Ca II K emission from HD 179949. The same
amount of energy flux can be also achieved for the case of
a dipole field for the hot jupiter ($p=2$, $q=3$) where
$a=0.045$~AU, $B_{*}$ = 250~G, and $B_p$ = 10~G.
If both the star and the planet have dipole fields
($p=q=3$), very strong fields $B_*=1000$ G and $B_p=30$ G are
required to generate the same energy flux.
Therefore the tight energy budget constrained by the synchronous
Ca II emission from HD 179949 strongly suggests that
the mean global fields of this F-type star are not likely
in a dipole field configuration at the location of the
planet but have the radial, open structure (i.e. $p=2$)
just like the solar fields have as a result of the outflowing
winds.

The argument
that hot jupiters might have weaker fields than our
Jupiter due to slower spin rates and weaker convection
(S\'anchez-Lavega 2004) should be treated with caution since
in addition to the uncertainty
in the interior structure of the metallic hydrogen
region of a hot jupiter,
the response of slow convection to
various rotation rates in the dynamo process is
not well understood.
If $p=2$, and $B_p \lesssim$ 1~G (S\'anchez-Lavega 2004)
for HD 179949 b, then eq(\ref{eq:energyflux}) indicates that
$B_{*}$ = 300~G is required to generate
the energy flux 10$^{5}$ erg~cm$^{-2}$~s$^{-1}$ and
in this case the stellar field dominates over the
planetary field even on the surface of the hot jupiter
(i.e. $q=\infty$ in eq(\ref{eq:energyflux})).
The observational
constraints on the strength of the stellar magnetic fields
such as the field versus Ca~II relation (Schrijver et al.~1989, 1992) along with
the radio cyclotron emissions from
the hot jupiter's magnetosphere\footnote{The characteristic frequency of the
cyclotron radiation is ${eB_p \over 2\pi m_e c}$,
where $c$ is the speed of light, $e$ and $m_e$ are the electron
charge and mass respectively. The radiation can reach $\approx 30-60$ MHz if $B_p \approx 10-20$ G.}
should help to narrow down the field strength of
hot jupiters in the magnetic interaction scenario,
therefore improving our knowledge of the interior structure
and the dynamo processes of gaseous planets.

Now we turn our attention to the variation of Ca II level, the phase
coverage of additional emission, and the phase lead at the
different epochs of our observations. The data for HD~179949
seem to suggest that the additional Ca~II emission is
smaller and the range of phase spanned by the emission is larger when the average (or minimal)
K emission is larger. Presumably this has something to do with the
intrinsic stellar activity and therefore the stellar field geometry.
The observed phase lead may be caused by spiral stellar
fields loaded with stellar winds.
When a star is in its quiet phase and therefore the average
Ca~II emission is very low, no enhancement
happens probably because $V_{\rm wind} \gtrsim V_A$ for a hot jupiter
located far out from its parent star and therefore
being outside the Alfv\'en radius, as may be the case
for the 2001 August observations of $\upsilon$ And.
As the star shifts to its more active phase,
the average Ca~II emission is at the moderate level and
$V_{\rm wind}$ is not far smaller than $V_{A}$ at
the hot jupiter. At this time, the configuration of
the field lines on the surface of the star may be
characterized by the open fields from the coronal holes
covering a large area of the star, as well as the closed
fields distributed only near the magnetic equator.
Consequently the foot-points of the open stellar field
lines pointing to the planet are located at lower latitudes,
as indicated by the 2001 and 2002 data for
HD 179949 fitted by a truncated sine curve
(Shkolnik et al. 2003; also shown in Figure~\ref{intK2}). While a great number
of closed fields form at low magnetic latitudes during
this time, the coronal holes shrink to the magnetic polar regions.
The bright spot at a high latitude implied from the observation
for HD 179949 in September 2003 might be caused by the scenario
that the hot jupiter perturbs the open field lines emanating from
the shrinking coronal holes near the polar regions, leading to a longer
phase duration of the additional Ca II flux and perhaps reducing the
additional emission due to the smaller projection along the line of
sight at higher latitudes. Note that
at the time close to the stellar-maximum activity,
the stellar field lines might be occasionally stretched out like
solar streamers emanating from the low latitudes of the star
where closed loops of stellar magnetic fields aggregate,
possibly giving rise to the planet-induced heating at low latitudes
of the stellar surface as well.

The picture that we have sketched
thus far assumes that most of
energy flux released from the vicinity of the hot jupiter is
transported along the field lines by Alfv\'en waves and deposited at the
foot-points of the magnetic lines. Since the field lines
inside the Alfv\'en radius are dominated by the poloidal
component, detailed calculations for stellar wind models
are needed to study how the integration of small pitch
angles of the field line can lead to
the moderate to large phase angles, $60^{\circ}$ for HD~179949 and
$180^{\circ}$ for $\upsilon$~And. The phase difference
is also determined by the Alfv\'en-speed travel time along
the stellar field line. If $V_{A}>>V_{\rm wind}$,
the Alfv\'en disturbance roughly takes
$a/V_A \approx$ a few
hours to propagate from the hot jupiter to the star with
$V_A \approx 10^7-10^8$ cm~s$^{-1}$ at 0.04 AU. This means that
the azimuthal angle that the planet has already traveled
over the Alfv\'en-speed travel time is not small.
Since the hot jupiter
of $\upsilon$~And is located farther out ($a=0.06$ AU) from its star than HD~179949 b ($a=0.045$ AU), the large phase
angle of $180^{\circ}$ for $\upsilon$~And might actually represent a phase
lag caused by the small inward group velocity of the
Alfv\'en waves $V_A-V_{r,{\rm wind}}$ that takes considerable
amount of time to travel along the field lines right
after the waves are generated from the planet.
Alternatively, a large phase
difference between the location of the heating spot
in the chromosphere and the position of the hot jupiter
might be caused by the entangled or rotationally spiraled magnetic fields
connecting directly between the hot jupiter and the star
(Rubenstein \& Schaefer 2000, Saar et al.~2004, Ip et al. 2004).

Besides considering the energy transported along the field lines and field
geometry, understanding how the energy is dissipated is important to
construct a complete picture for the Ca~II emission in the planet-induced
scenario. In our own solar system, a bow shock of the solar wind generally
hinders this because Alfv\'en waves cannot propagate toward the star from
the region outside the Alfv\'en radius.
The inward Poynting energy of Alfv\'en waves $r^2 b^2 B/\sqrt{\rho}$ is
roughly conserved if the wind velocity $V_{r,{\rm wind}}$ is much smaller
than the Alfv\'en speed $V_A$. Here $r$ is the distance away from the
star, $b$ is the magnetic disturbance of the wave, and $\rho$ is the mass
density of the stellar wind. At first glance, the transition region
characterized by a steep decrease of the Alfv\'en speed with depth seems
to provide a possible radial stratification for the inward-propagating
Alfv\'en waves to pile up the magnetic energy density, leading to the
nonlinear dissipation of the growing waves
and therefore the heating on the top of the chromosphere.
However, the extremely narrow transition region corresponding to
sharp gradient in density and Alfv\'en speed should act as a wave barrier
to reflect the Alfv\'en waves. In this case, the wavelength is
comparable to the size of the magnetopause of the hot jupiter
(Wright 1987) unless the high-frequency modes are
largely excited at the reconnection site. The energy carried by the planet-induced Alfv\'en waves along
the open fields might be finally transmitted to coronal loops
and therefore might be dissipated via resonance absorption,
a damping mechanism however in contradiction to the observations
of the solar corona and the coronae of cool stars
(Schrijver \& Aschwanden 2002; Demoulin et al.~2003).
The heating due to accelerated particles by Alfv\'en waves
(Crary 1997) is not important either because unlike the
interplanetary space, the dense corona eliminates the kinetic
effect of plasmas inside 0.04 AU.\footnote{Besides the particle
acceleration by Alfv\'en waves, a jet of accelerated particles
with high speeds from the reconnection site at the magnetopause
of a magnetized planet might be able to heat the stellar corona,
but it is difficult for the process to produce a large phase
difference between the heating spot and the location of the
planet unless the coronal fields can be entangled in a more
large-scale manner like the models for the RS CVn stars.}
Despite the difficulties, the energy deposit on the surface of
the star from the planet-induced Alfv\'en disturbances may be
achieved by interacting non-linearly with stellar winds. The
stellar wind consists of charges particles, stellar fields,
and probably Alfv\'en waves. Non-linear interactions between
planet-induced incoming and stellar intrinsic outgoing
waves is one route of heating the surface of the star.

The scenario of magnetic interaction implies the orbital decay of
hot jupiters
since the ultimate source of energy comes from the orbital energy.
Theoretically the orbital decay of hot jupiters on the timescale of
a few billion years can result from the tidal
dissipation in the host star driven by the hot jupiter so long as
the tidal dissipation in the solar-type stars is efficient
(Rasio et al.~1996, Witte \& Savonije 2002, P\"atzold \& Rauer 2002, Jiang et al.~2003).
In the magnetic interaction scenario
for Ca II emission, the time scale
of orbital decay is roughly equal to the ratio of the orbital energy of the
hot jupiter ($\sim$10$^{44}$ ergs) to $10^{27}$ erg/s. This gives a timescale
as short as several billion years, imposing a non-negligible constraint on
modeling the orbital evolution of hot jupiters.

In the case of $\tau$ Boo, the F-type star might be almost tidally locked to
its hot jupiter. Therefore there is less free energy available from the
planet's orbit. According to eq(\ref{eq:energyflux}),
the azimuthal relative motion between the stellar and the planetary
magnetospheres is smaller than that for
the HD 179949 system by a factor of 0.2 if 3.2 d is indeed the rotation
period of $\tau$ Boo. The mean chromospheric emission $\langle$K$\arcmin\rangle$ from $\tau$ Boo is
weaker than that from HD 179949 roughly by a factor 0.9. Assuming
that the mean stellar field of $\tau$ Boo is smaller than that
of HD 179949 by 0.9$^2$ ($\langle$K$\arcmin\rangle$ $\propto$ $B_{*}^{0.5}$), one can estimate that the input
energy to $\tau$ Boo is less than that to HD 179949 roughly by a
factor of 0.15. If the corresponding Ca~II emissions are dimmed by
this same factor, this effect should have been detected. However unlike
the other two F-type stars HD 179949 and $\upsilon$~And, the variability
of Ca~II emissions from $\tau$ Boo did not show any consistent phase relation
with the planet's orbit. Note that the radial movement of
spiral stellar fields with the stellar winds might be important in providing
energy in this case because $V_{r,{\rm wind}} \approx V_{orb}$ at 0.04 AU in
the solar system. However in the case of small pitch angles of the field
lines inside the Alfv\'en radius, the Alfv\'en modes driven solely by the
radial impact of the stellar winds might not be as efficient as the stellar
fields sweeps across the hot jupiter due to the slow relative azimuthal motion.

The transiting system HD 209458 did not show synchronized Ca II
enhancement, perhaps because G stars have $V_{r,{\rm wind}} > V_A$
at the distance of the hot jupiter due to weaker stellar fields.
The non-synchronized Ca II enhancement of $\tau$ Boo and HD~209458 (as well
as the flaring on HD 73256 and $\kappa^{1}$~Ceti, discussed in Section~\ref{modbyrot})
might still be attributed to direct interactions with their planets but
through more chaotic means due to the variability of the
stellar winds at the locations of
these hot jupiters relative to the nearby Alfv\'en radius.
Monitoring these stars continuously through
several orbital and rotational periods should pin down the
cause $-$ intrinsic due to fast rotation or induced by a
hot jupiter $-$ of the strong night-to-night variabilities
detected in these systems.

\section{Summary and Future}\label{summary}

Of the sample of stars observed from the CFHT, those with planets (with the exception of 51~Peg) show significant night-to-night variations in their Ca~II H and K reversals. $\tau$~Ceti and the Sun, which have no close planets, remained very steady throughout each of the four observing runs. HD~179949 and $\upsilon$~And exhibited repeated orbital phase-dependent activity with enhanced emission leading the sub-planetary point by 0.17 and 0.47 in orbital phase, respectively. Both systems are consistent with a magnetic heating scenario and may be the first glimpse at the magnetospheres of extrasolar planets. The phase-lead or lag of the peak emission relative to the sub-planetary longitude can provide information on the field geometries and the nature of the effect such as tidal friction, magnetic drag or reconnection with off-center magnetic fields, including a Parker-spiral type scenario.

$\tau$ Boo and HD 209458 also exhibited night-to-night variations that could not exclusively be due to stellar rotation. If $\tau$ Boo is indeed tidally locked by its hot jupiter, there is no orbital energy available to generate the Alfv\'en modes efficiently as the stellar fields sweep across the planet due to the slow relative azimuthal motion. We measured the excess absolute flux released in the enhanced chromospheric emission of HD~179949 to be the same order of magnitude as a typical solar flare, $\sim$~10$^{27}$ erg~s$^{-1}$ or 1.5$\times$10$^{5}$ ergs~s$^{-1}$~cm$^{-2}$. This implies that flare-like activity triggered by the interaction of a star with its hot jupiter may be an important energy source in the stellar outer atmosphere. This offers a mechanism for short-term chromospheric activity.

The H \& K emission of $\kappa^{1}$~Ceti, an active star with no confirmed planet, was clearly modulated by the star's 9.3-d rotation.  Similarly, HD 73256 displayed rotational modulation with its 14-d period.  In these two cases, the chromospheric emission increases by $\approx$~6\% (relative to the normalization level at 1/3 of the continuum). Any planet-induced heating at the level of 1 $-$ 2\% could have been diluted by the dominating hotspot on the stellar surface.  Neither the $\Delta$RVs nor the Ca II periodicity exclude the possibility of a sub-stellar companion in a tight orbit around $\kappa^{1}$~Ceti.

Apart from the cyclical component for four of the stars, short-term chromospheric activity appears weakly dependent on the mean K-line reversal intensities for the sample of 13 stars. Also, a suggestive correlation exists with $M_{p}$sin$i$ and thus with the hot jupiter's magnetic field strength. Because of their small separation ($\leq$ 0.1
AU), many of the hot jupiters lie within the Alfv\'en radius of their host stars which allows a direct magnetic interaction with the stellar surface.

Additional Ca II observations are crucial to confirm the stability of the magnetic interaction as well as to establish better phase coverage. Observations on timescales of a few years will begin to characterize the long-term activity of our program stars and allow us to see correlations between intrinsic Ca II
emission and night-to-night activity more clearly. This work
opens up the possibility of characterizing
planet-star interactions with implications for extrasolar planet
magnetic fields and the energy contribution to stellar atmospheres.

A next step in understanding planet-star interactions is to map the activity as a function of stellar atmospheric height. Above the chromosphere lies the thin transition region (TR), where the temperature increases steeply as density and pressure drop, and the corona, which can extend out to several stellar radii. And since the magnetic field drops off as $r^{-p}$ (where $2 \leq p \leq 3$), these layers facilitate a stronger interaction with the planet. Their FUV and X-ray emissions will be extremely important diagnostics.  One indication that the heating is from the outside in is if the increase in emission occurs slightly earlier in phase than in Ca II. Moreover, the relative strengths
of the different emission lines will tell us where most of the energy is
dissipated. The energy sum will point out if
there are any discrepancies with the theorized energy budget. Orbital phase-dependent variability at these heights will constrain further the nature, form and strength of the interaction as well as specify non-thermal radiative processes in these hot layers of gas.

\acknowledgements

We are grateful to Marek Wolf and Petr Harmanec for their photometric observations of HD 179949 made at the South African Astronomical Observatory (SAAO). We thank Geoff Bryden, Peng-Fei Chen, Gary Glatzmaier, Gordon I. Ogilvie, and Ethan T. Vishniac for useful communications regarding Section~\ref{theory}. Research funding from the Canadian Natural Sciences and Engineering Research Council (G.A.H.W. \& E.S.) and the National Research Council of Canada (D.A.B.) is gratefully acknowledged. We are indebted to the CFHT staff for their care in setting up the CAFE fiber feed and the Gecko spectrograph, as well as to the staff at ESO's VLT for their telescope and instrument support and the real-time data-reduction pipeline. Also, we appreciate the helpful comments and suggestions from the referee, Steve Saar.

\clearpage

\end{document}